\documentclass[nofootinbib,reprint,amsmath,amssymb,aps]{revtex4-1}
\usepackage{graphicx}
\usepackage[utf8,latin1]{inputenc}
\usepackage{dcolumn}
\usepackage{bm}
\usepackage{hyperref}
\usepackage{amsmath}
\newcommand{\qm}[1]{``#1''}
\usepackage{hyperref}
\hypersetup{colorlinks, linkcolor={red},citecolor={blue},urlcolor={blue}}

\begin{document}

\preprint{APS/123-QED}

\title[Dissipative systems in metric theories of gravity. Foundations and applications of the energy formalism]{Dissipative systems in metric theories of gravity.\\ Foundations and applications of the energy formalism}

\author{Vittorio De Falco$^{1}$},\email{vittorio.defalco@physics.cz}
\author{Emmanuele Battista$^{1,2,3}$\vspace{0.5cm}}\email{emmanuelebattista@gmail.com}

\affiliation{$^1$ Research Centre for Computational Physics and Data Processing, Faculty of Philosophy \& Science, Silesian University in Opava, Bezru\v{c}ovo n\'am.~13, CZ-746\,01 Opava, Czech Republic\\
$^2$ Universit\`{a} degli studi di Napoli \qm{Federico II}, Dipartimento di Fisica \qm{Ettore Pancini}, Complesso Universitario di Monte S. Angelo, Via Cintia Edificio 6, 80126 Napoli, Italy\\
$^3$ Istituto Nazionale di Fisica Nucleare, Sezione di Napoli, Complesso Universitario di Monte S. Angelo, Via Cintia Edificio 6, 80126 Napoli, Italy
}

\date{\today}

\begin{abstract}
In this paper we introduce a new procedure, termed by us \emph{energy formalism}, to deal with dissipative systems in metric theories of gravity. This approach aims at determining the analytic expression of Rayleigh dissipation function in the context of the inverse problem in the calculus of variations. We describe our method in detail, presenting a simple example. After, we consider as first extensive application the general relativistic Poynting-Robertson effect. The obtained results and future implications are discussed.
\end{abstract}



\maketitle

\section{Introduction}
\label{sec:intro}
Dissipation is a subject which concerns several research fields, ranging from classical to quantum physics. Apart from the usual picture regarding the waste of stored (mechanical) energy undergone by a dynamical system during time evolution, its meaning has to be understood in a broader sense. For instance, dissipation of entropy in the framework of kinetic theory of rarefied gases \cite{Carrillo2001}, dissipative phenomena in terms of decoherence effects at quantum level \cite{Breuer2002,Nielsen2000,Alicki2018,Caldeira1981}, dissipative colliding particles to model warm or hot dark matter in cosmological settings \cite{Moore1994}, and lastly, as a more \qm{exotic} situation, dissipation of information in the realm of string theory \cite{Hooft1999}.

Dissipation configures as a fundamental ingredient to make a model more realistic, although the mathematical framework spreads more and more out of control. Indeed, dissipative systems, albeit widely explored in the literature, present some  critical consequences, like: loss of existence, smoothness, and symmetries of the original solution (e.g., the unsolved Millennium Prize Problem of Navier-Stokes equations \cite{Tao2016}), presence of topologically complex structures featuring chaotic behaviour (e.g., Sinai-Ruelle-Bowen measure, which has been awarded the 2014 Abel Prize \cite{Sinai1972,Bowen1975,Ruelle1976,Sinai2017}), production of quasi-normal mode frequencies (related to, e.g., the recent Nobel Prize discovery of gravitational waves \cite{Berti2009,Cardoso2018}). 

Conventionally, time-independent dissipative forces can be formally expressed by 
\begin{equation} \label{eq:dissipation}
F_i(\boldsymbol{q},\boldsymbol{\dot q})=-\Phi_i(\boldsymbol{q},\boldsymbol{\dot q}),\quad i=1,\dots,N\ ,    
\end{equation}
where $\Phi_i$ are non-negative functions usually expressed as $\Phi_i(\boldsymbol{q},\boldsymbol{\dot q})=g_i(\boldsymbol{q})\Psi_i(\boldsymbol{\dot q})$,  $g_i$ and $\Psi_i$ being non-negative functions. Classically, the most common functional forms for $\Psi_i$ are represented by: power or polynomial functions, i.e., $\Phi_i(\boldsymbol{\dot q})=\left( {\dot q}_i\right)^n$, where  depending on the values of the exponent we have frictional ($n=0$), viscous ($n=1$), and high-velocity frictional ($n\ge2$) forces; logarithmic functions, i.e., $\Phi_i(\boldsymbol{\dot q})=\lg{\dot q}_i$; other kinds of elementary functions obtained by fitting observational data \cite{Braun2011,Manini2016}. In metric theories of gravity, such forces strongly couple with the geometrical structure of spacetime, giving rise to non-linear functions $\Phi_i$ \cite{Ridgely2010}. 

The equations of motion of dynamical systems may be derived by the principle of least action through the Euler-Lagrange equations. If we regard the Lagrangian as the unknown function, we are led to the renowned inverse problem of the calculus of variations \cite{Santilli1978,Morandi1990,Do2016}. Matters complicate considerably when dissipation phenomena are taken into account. Indeed, the analysis to capture the notion of a \emph{well-posed problem} (in the sense of Hadamard) becomes a subtle task (see Refs. \cite{Gitman2007b,Mestdag2011,Gitman2007,Kochan2010,Kabanikhin2011} and references therein).

The strong efforts made over the course of time to achieve well-posedness represent the safe-pass to look for solutions. The approaches aimed at investigating problems involving dissipation constitute a huge domain. To give a clear idea about the state of the art, we classify such methods in three categories:

\begin{itemize}
\item \emph{quantitative} frameworks, represented by mathematical analytical techniques. These involve either general classical solutions or more advanced methods dealing with weak solutions  \cite{Evans2010,Brezis2010,Federer2014};

\item \emph{qualitative} patterns, relying on dynamical systems theory. The modern techniques are based on chaos theory, which still constitutes an active research field, although there are aspects not yet fully understood \cite{Sprott2011,Li2016}.

\item \emph{numerical} approaches, constituting at present a more and more valid support. The actual increasing computational power and the recent development of advanced numerical methods \cite{Quarteroni2009,Quarteroni2015} permit to describe accurately several complex dissipative systems, but they require both a major international effort in producing the massive codes and the necessity to have supercomputers for the prolonged computational times.

\end{itemize}

Even though the abovementioned frameworks have allowed several classical thorny issues to be successfully addressed in their deeper aspects, 
many of them turn out to be not suited enough to work out numerous challenging issues within the context of metric theories of gravity (e.g., magnetohydrodynamical/hydrodynamical simulations in high-energy astrophysics \cite{Rezzolla2013}, dissipative collapse of axially symmetric sources \cite{Herrera2014}). Due to these subtleties, nowadays there exists a widespread attitude to approach difficult problems by entrusting them mainly through appropriate numerical codes. However, despite being a precious resource, these sometimes discourage theoretical investigations. 

A natural consequence is that analytic procedures, which are fundamental to have direct insight into mathematical and physical details of the model under investigation, are becoming more and more rare. Therefore, the introduction of analytic techniques might be extremely valuable. It is in that spirit that we developed a \qm{formalism} to analytically determine the Rayleigh potential of the general relativistic Poynting-Robertson (PR) effect. However, we will show how our new method, which we term \emph{energy formalism}, can be applied to a variety of different problems involving dissipation. 


The Rayleigh dissipation function represents a valuable tool, because it permits to: $(i)$ solve a broader class of problems with respect to the conservative ones \cite{Santilli1978}; $(ii)$ completely describe the (velocity dependent) dissipative systems \cite{ Goldstein2002} without resorting to unconventional approaches, which alter the physical interpretation of the functions characterising the dynamics (see e.g., Ref. \cite{Galley2013}); $(iii)$ reconstruct the energy dissipated by the system \cite{Whittaker1988}; $(iv)$ perform the analysis of the stability of the equilibrium configurations through Lyapunov theory \cite{Pars1965}.
In addition, in this paper we will show new implications of the Rayleigh potential in the study of dissipative systems.

The article is organised as follows: in Sect. \ref{sec:EF} we introduce the fundamental aspects of the \emph{energy formalism}; in Sect. \ref{sec:TPE} we consider it in a test example in classical mechanics; in Sect. \ref{sec:AEF} we apply it to the general relativistic PR effect;  in Sec. \ref{sec:end} we discuss our results and comment on future applications. 

We note that in a separate letter \cite{DBletter2019} we have presented a brief and basic account of our new formalism and of the main results we obtained. In this paper, we aim at giving a more detailed analysis of the method in its formal aspects and applications, and at generalising its validity in generic metric theories of gravity.



\section{Energy formalism}
\label{sec:EF}
\emph{Energy formalism} is very useful for the determination of the primitive of an exact differential semi-basic one-form (i.e., the Rayleigh potential in physics-oriented literature) in models where dissipative effects occur. Typical examples of analytic Rayleigh functions discussed in the literature involve simple (mostly classical) models where the dynamical equations can be easily integrated yielding a quadratic function in the velocities \cite{Goldstein2002}. Therefore, the introduction of innovative techniques, able to encompass new features in gravity patterns, becomes a valuable tool. Our method, being collocated in the context of potential theory of dissipative systems, might open up a new research area in the field of inverse problems.

In the following sections, we introduce the formal aspects of the \emph{energy formalism} in a general framework. We establish the necessary hypotheses under which it can be successfully employed. 
The geometrical setup underlying the model imparts it such a general structure that it can be implemented in any metric theory of gravity.



\subsection{Geometrical setting}
\label{sec:geo_sett}
Let $\mathcal{M}$ be an $n$-dimensional ($n$ being a finite integer such that $n\ge1$), real, topological, differential, pseudo Riemannian, asymptotically flat, and simply connected manifold \cite{Neill1983,Joshi1983,Nakahara2003,Lee2010} endowed with:
\begin{itemize}
\item a Hausdorff topology $\mathcal{T}$, whose elements are the open sets in $\mathcal{M}$ \cite{Abraham1978,Nakahara2003}; 
\item an atlas $\mathcal{A}=\left\{(\mathcal{U}_\alpha,\varphi_\alpha)\mid\alpha \in \mathcal{I}\right\}$ indexed by a set $\mathcal{I}$ of charts $(\mathcal{U}_\alpha,\varphi_\alpha)$ on $\mathcal{M}$. Each small patch of the manifold $\mathcal{M}$ can be labelled by a set of local coordinates $\boldsymbol{X}=(X^1,...,X^n)$ on $\mathcal{M}$ \cite{Lee2010};
\item a non-degenerate, differentiable, symmetric, bilinear metric tensor $\mathfrak{g}:T_p\mathcal{M}\times T_p\mathcal{M}\to \mathbb{R}$, $T_p \mathcal{M}$ being the tangent space to $\mathcal{M}$ at a point $p \in \mathcal{M}$ \cite{Misner1973};
\item the induced Lebesgue measure $\mathfrak{m}:\Sigma\to[0,+\infty]$, where $\Sigma$ is a $\sigma$-algebra defined over $\mathcal{M}$ \cite{Rudin1987}. 
\end{itemize}

\subsubsection{Tangent and cotangent bundle}
\label{sec:tan_cotan_bund}
Let $T\mathcal{M}$ denote the tangent bundle of $\mathcal{M}$, defined as \cite{Abraham1978,Nakahara2003}
\begin{equation}
T\mathcal{M}=\bigcup_{p\in\mathcal{M}} T_{p}\mathcal{M}\equiv \left\{(p,\boldsymbol{v})| p\in \mathcal{M}, \boldsymbol{v}\in T_{p}\mathcal{M}\right\}.
\end{equation}

It can be considered as a $2n$-dimensional manifold endowed with an induced locally Hausdorff topology and an atlas from $\mathcal{M}$. 
This permits to define a real and continuous manifold structure on $T\mathcal{M}$ 
allowing every neighbourhood of a point $(p,\boldsymbol{v})\in$ $T\mathcal{M}$ to be labelled by a set of local coordinates $(\boldsymbol{X},\boldsymbol{U})$. 
Furthermore, $T\mathcal{M}$ is a simply connected domain, because $\mathcal{M}$ is simply connected by hypothesis, whereas $T_p\mathcal{M}$ is simply connected for all $p\in\mathcal{M}$, since it is a vector space \cite{Lee2010}. 

The dual space of $T\mathcal{M}$ is indicated with $T^*\mathcal{M}$ and it is referred to as the cotangent bundle of $\mathcal{M}$, defined as 
\begin{equation}
T^*\mathcal{M}=\bigcup_{p\in\mathcal{M}} T^*_{p}\mathcal{M}\equiv \left\{(p,\boldsymbol{f})| p\in \mathcal{M}, \boldsymbol{f}: T_{p}\mathcal{M}\to \mathbb{R}\right\},
\end{equation}
where $T^*_{p}\mathcal{M}$ is the cotangent space of $\mathcal{M}$ at the point $p\in\mathcal{M}$ \cite{Abraham1978,Nakahara2003}. 
$T^*\mathcal{M}$ configures as a $2n$-dimensional manifold endowed with an induced topology and an atlas. 

\subsection{Exact differential semi-basic one-form}
\label{sec:ex_form}
Let $\boldsymbol{\omega}:T\mathcal{M}\rightarrow T^*\mathcal{M}$ be a smooth, differential semi-basic one-form \cite{Libermann1987} (also known in the literature as \emph{one-form along the tangent bundle projection} \cite{Martinez1992,MARTINEZ19931,Mestdag2011}), where \emph{smooth} indicates, throughout the paper, a function of class $\mathcal{C}^k(T\mathcal{M},\mathfrak{m})$ with $k\ge1$, i.e., $\mathfrak{m}$-continuous with the first $k$ derivatives $\mathfrak{m}$-continuous. 

Since $\boldsymbol{\omega}$ is defined on $T\mathcal{M}$, it can be written, in the local coordinates $(\boldsymbol{X},\boldsymbol{U})$ of $T\mathcal{M}$, as
\begin{equation} \label{differential_oneform_1}
\boldsymbol{\omega}(\boldsymbol{X},\boldsymbol{U})=F^\alpha(\boldsymbol{X},\boldsymbol{U})\ \boldsymbol{{\rm d}}X_\alpha,
\end{equation}
where $F^\alpha(\boldsymbol{X},\boldsymbol{U})$ are referred to as the $\mathcal{C}^k(T\mathcal{M},\mathfrak{m})$ components of the differential semi-basic one-form $\boldsymbol{\omega}(\boldsymbol{X},\boldsymbol{U})$. From the above equation, it is clear that a semi-basic one-form belongs to a particular class of forms defined on $T\mathcal{M}$ having smooth components depending on both $\boldsymbol{X}$ and $\boldsymbol{U}$.


In addition, we assume that $\boldsymbol{\omega}$ is closed under the \emph{vertical exterior derivative} $\boldsymbol{{\rm d^V}}$ \cite{Abraham1978,MARTINEZ19931,Mestdag2011}, i.e., $\boldsymbol{{\rm d^V}}\boldsymbol{\omega} = 0$. The local expression of this operator is given by
\begin{equation} \label{eq:vertical_derivative1}
\boldsymbol{{\rm d^V}}F= \frac{\partial F}{\partial U_\alpha} \boldsymbol{{\rm d}}X_\alpha, \quad \forall F\in \mathcal{C}^k(T\mathcal{M},\mathfrak{m}),\ k\ge1.
\end{equation}
Therefore, the closure condition $\boldsymbol{{\rm d^V}}\boldsymbol{\omega}=0$ implies the equality among the cross derivatives, i.e., $\partial F^\alpha/ \partial U_\beta=\partial F^\beta/ \partial U_\alpha$. The Poincar\'e lemma, adapted to the case of vertical differentiation \citep{Martinez1992}, guarantees that $\boldsymbol{\omega}$ is also exact, i.e., it can be expressed as the vertical exterior derivative of a 0-form $V(\boldsymbol{X},\boldsymbol{U})\in \mathcal{C}^k(T\mathcal{M},\mathfrak{m})$, namely
\begin{equation} \label{eq: differential_of_V}
-\boldsymbol{{\rm d^V}} V(\boldsymbol{X},\boldsymbol{U})\equiv
-\frac{\partial V(\boldsymbol{X},\boldsymbol{U})}{\partial U_\alpha}\boldsymbol{{\rm d}}X_\alpha=\boldsymbol{\omega}(\boldsymbol{X},\boldsymbol{U}),
\end{equation}
where $V$ is called primitive (or potential function, in the physics-oriented literature) of $\boldsymbol{\omega}$. 

\subsection{Energy function}
\label{sec:en_coinst}
We consider the following \emph{energy function}\footnote{\label{footnote}We note that Eq. (\ref{eq:constraint}) does not represent the most general form that the \emph{energy function} may assume. Indeed, it might occur to have a smooth implicit function $\chi:\mathbb{R}^{2n+1}\to\mathbb{R}$ such that $\chi(\boldsymbol{X},\boldsymbol{U},\mathbb{E})=0$. If, for each point $(\boldsymbol{X}_0,\boldsymbol{U}_0,\mathbb{E}_0)\in \mathbb{R}^{2n+1}$ such that $\chi(\boldsymbol{X}_0,\boldsymbol{U}_0,\mathbb{E}_0)=0$, we have $\frac{\partial \chi}{\partial \mathbb{E}}(\boldsymbol{X}_0,\boldsymbol{U}_0,\mathbb{E}_0)\neq0$, then for the \emph{implicit function theorem} there exists an open set $\mathcal{U}\times\mathcal{D}\subseteq \mathbb{R}^{2n}$ containing $(\boldsymbol{X}_0,\boldsymbol{U}_0)$ and an open interval $\mathcal{V}\subseteq \mathbb{R}$ containing $\mathbb{E}_0$ such that exists and it is unique a smooth function $\rho:\mathcal{U}\times\mathcal{D}\to\mathcal{V}$ such that $\chi(\boldsymbol{X}_0,\rho(\boldsymbol{X}_0,\boldsymbol{U}_0))=0$ and $\mathbb{E}_0=\rho(\boldsymbol{X}_0,\boldsymbol{U}_0)$ \cite{Munkres1997}, obtaining thus Eq. (\ref{eq:constraint}). Instead, if $\frac{\partial \chi}{\partial \mathbb{E}}(\boldsymbol{X}_0,\boldsymbol{U}_0,\mathbb{E}_0)=0$, the hypothesis of the implicit function theorem is not satisfied anymore, therefore we have to rely only on numerical methods (also called \emph{root-finding algorithm}) to determine the value of $\mathbb{E}$ in terms of $(\boldsymbol{X},\boldsymbol{U})$ (see Chap. 9 in Ref. \cite{Press2002} for further details).}:
\begin{equation} \label{eq:constraint}
\mathbb{E}=\rho(\boldsymbol{X},\boldsymbol{U}),
\end{equation}
where $\rho:T\mathcal{M}\to\mathbb{R}$ is a smooth function of the local coordinates $(\boldsymbol{X},\boldsymbol{U})$ of $T\mathcal{M}$,
and $\mathbb{E}$ the physical quantity dissipated by the system. For simplicity, $\mathbb{E}$ is referred to as energy. However, this method is general in such a way that it can be applied to different dissipative processes. Therefore, $\mathbb{E}$ can represent any physical parameter associated to the system under
investigation (e.g., orbital angular momentum, mass, spin, and so on). It is important to note that $\mathbb{E}$ represents a way to label what is reported at the right-hand side of Eq. (\ref{eq:constraint}). Therefore, $\rho(\boldsymbol{X},\boldsymbol{U})$ is the operative definition of $\mathbb{E}$, i.e., all the manipulations and calculations regarding $\mathbb{E}$ transfer operationally on $\rho$. We have stressed this point, even if it may appear trivial, since it is crucial for the overall comprehension of the formalism and the following calculations. Moreover, it should be noted that Eq. (\ref{eq:constraint}) defines a $(2n-1)$-dimensional hypersurface, that we term \emph{energy hypersurface}, embedded in the $2n$-dimensional space spanned by the local coordinates system $(\boldsymbol{X},\boldsymbol{U})$ of $T\mathcal{M}$. This hypersurface changes at each proper time instant $\tau$, i.e., we have $\mathbb{E}=\mathbb{E}(\tau)$.

\subsection{Energy operator}
\label{sec:en_op}
The condition (\ref{eq:constraint}) allows us to consider the components of the differential semi-basic one-form (\ref{differential_oneform_1}) as a function of the local coordinates and the energy, i.e.,
\begin{equation} \label{eq:constraint_2}
F^\alpha=F^\alpha(\mathbb{E},\boldsymbol{X},\boldsymbol{U}).
\end{equation}
This equation can be obtained in different ways. In the most manageable situation, one can simply replace $\rho$ with $\mathbb{E}$ each time the former occurs in the expression of $F^\alpha$. Otherwise, in the most general case, one can exploit (\ref{eq:constraint}) to express one coordinate $U^\alpha$ in terms of the remaining $U^\beta$ (with $\beta\neq\alpha$), of the local coordinates $\boldsymbol{X}$, and of the function $\mathbb{E}$ through either the implicit function theorem or the root-finding algorithms to reconstruct the inverse function (see footnote \ref{footnote}, for further details). This process does not alter the number of independent variables, because the variable $U^\alpha$ has been replaced by $\mathbb{E}$.

Equations (\ref{eq:constraint}) and (\ref{eq:constraint_2}) represent the key aspects of the \emph{energy formalism}. In fact, as we will see in the next sections, this approach turns out to be a powerful method for the research of the potential function $V(\boldsymbol{X},\boldsymbol{U})$, since it allows us to simplify cumbersome calculations. 

From the above considerations, it is clear that we can consider the derivative operator in terms of the energy function through the chain rule, i.e.,
\begin{equation} \label{eq:trader}
\frac{\partial(\ \cdot\ ) }{\partial U_\alpha}=\frac{\partial  \rho(\boldsymbol{X},\boldsymbol{U})}{\partial U_\alpha}\frac{\partial (\ \cdot\ )}{\partial \mathbb{E}}.
\end{equation}

\subsection{Primitive in terms of energy}
\label{sec:prim_en}
The primitive function $V(\boldsymbol{X},\boldsymbol{U})$ satisfies the condition
\begin{equation}\label{eq:primitive_original}
-\frac{\partial V}{\partial U_\alpha}=F^\alpha,
\end{equation}
which, once Eq. (\ref{eq:trader}) has been employed, can be recasted as
\begin{equation} \label{eq:primitive}
-\frac{\partial \rho}{\partial U_\alpha}\frac{\partial V}{\partial \mathbb{E}}=F^\alpha.
\end{equation}
Taking the scalar product of both members of Eq. (\ref{eq:primitive}) by a function $B(\boldsymbol{X},\boldsymbol{U},\mathbb{E})_\alpha$, chosen opportunely to simplify the calculations, we obtain
\begin{equation} \label{eq:mult}
-\frac{\partial \rho}{\partial U_\alpha}\frac{\partial V}{\partial \mathbb{E}} B(\boldsymbol{X},\boldsymbol{U},\mathbb{E})_\alpha = F^\alpha B(\boldsymbol{X},\boldsymbol{U},\mathbb{E})_\alpha. 
\end{equation}

Equation (\ref{eq:mult}) yields a differential equation for $V(\boldsymbol{X},\boldsymbol{U})$ in terms of $\mathbb{E}$, i.e.,
\begin{equation} \label{eq:solV} 
-\frac{\partial V}{\partial \mathbb{E}}=\frac{ F^\alpha B(\boldsymbol{X},\boldsymbol{U},\mathbb{E})_\alpha}{\frac{\partial \rho}{\partial U_\alpha}B(\boldsymbol{X},\boldsymbol{U},\mathbb{E})_\alpha}.
\end{equation}
The potential $V$ is thus given by
\begin{equation} \label{eq:pot_E}
V=\int \left(-\frac{F^\alpha B(\boldsymbol{X},\boldsymbol{U},\mathbb{E})_\alpha}{\frac{\partial \rho}{\partial U_\alpha}B(\boldsymbol{X},\boldsymbol{U},\mathbb{E})_\alpha} \right){\rm d}\mathbb{E}+f(\boldsymbol{X},\boldsymbol{U}),
\end{equation}
where $f(\boldsymbol{X},\boldsymbol{U})$ is a function of the local coordinates which is constant with respect to $\mathbb{E}$, i.e.,
\begin{equation} \label{eq:derivata_di_f}
\dfrac{\partial f(\boldsymbol{X},\boldsymbol{U})}{\partial \mathbb{E}}=0.
\end{equation}
The role of the function $f(\boldsymbol{X},\boldsymbol{U})$ is crucial in our \emph{energy formalism}, since it allows us to transfer all our ignorance regarding the definitive form of the potential $V$ to a function having vanishing derivatives with respect to the energy variable. Therefore, in this way whenever the research of the analytic form of the potential $V$ in terms of the local coordinates $(\boldsymbol{X},\boldsymbol{U})$ becomes unfeasible, we can at least set down an expression for the primitive as a function of $\mathbb{E}$ and hence describe the underlying dynamics by employing such a physical variable. 
The $f(\boldsymbol{X},\boldsymbol{U})$ function can be determined by applying the iterative process usually employed to integrate exact differential one-forms.

\subsection{Advantages of the energy formalism}
\label{sec:summary}
 
In summary, the \emph{energy formalism} assures the following advantages:
\begin{itemize}
    \item[($i$)] by exploiting the \emph{energy function} (\ref{eq:constraint}), it permits to simplify the calculations regarding the determination of the $V$ potential related to an exact differential semi-basic one-form $\boldsymbol{\omega}$ (cf. Eq. (\ref{eq: differential_of_V})) by expressing the force in terms of the energy variable $\mathbb{E}$, as shown in Eq. (\ref{eq:constraint_2}). This represent one of the key aspects of this method, because it reduces tremendously the calculations, passing from an integration involving the $n$ variables $\boldsymbol{U}$ to only one, represented by the energy $\mathbb{E}$ (see Eq. (\ref{eq:pot_E}));
    \item[($ii$)] in all those cases in which the determination of the $f(\boldsymbol{X},\boldsymbol{U})$ function defined in (\ref{eq:derivata_di_f}) still remains complicated, one can at least obtain an expression for the $V$ potential in terms of the $\mathbb{E}$ energy, as explained after Eqs. (\ref{eq:pot_E}) and (\ref{eq:derivata_di_f});
    
    \item[($iii$)] it represents a convenient approach to integrate the primitive of an exact differential semi-basic one-form for dissipative systems which can be broadly employed both in theoretical and applied physics, and even in pure mathematical analysis for its general geometrical presentation;
    \item[($iv$)] although our method has been pursued in the context of differential semi-basic one-forms, it can also be applied to differential one-forms $\boldsymbol{\alpha}:\mathcal{M}\to T^*\mathcal{M}$ defined over the simply connected manifold $\mathcal{M}$. In this case, all calculations performed in terms of $\boldsymbol{U}$ will be recasted as functions of the local coordinates system $\boldsymbol{X}$ only. Indeed, the vertical exterior derivative occurring in Eqs. (\ref{eq:vertical_derivative1})--(\ref{eq: differential_of_V}) will be replaced with the usual exterior derivative operator $\boldsymbol{{\rm d}}$ mapping $r$-forms on $\mathcal{M}$ in $(r+1)$-forms on $\mathcal{M}$ \cite{Abraham1978,Lee2010}. Furthermore, the energy function will be simply given by $\mathbb{E}=\rho(\boldsymbol{X})$. 
    
\item[($v$)] this formalism can be extended in general also to \qm{lifted} differential one-forms $\boldsymbol{\beta}:\mathcal{N}\to T^*\mathcal{M}$, where $\mathcal{N}$ can be identified with $T^m\mathcal{M}$ ($m$ being an integer such that $m\ge1$). $T^m\mathcal{M}$ represents an $[(m+1)n]$-dimensional smooth manifold defined via repeated application of the tangent bundle construction and it is known in the literature as the \emph{$m$-th-order tangent bundle} \cite{Lee2010}. Similarly to the case of $T\mathcal{M}$ outlined in Sec. \ref{sec:tan_cotan_bund}, it is possible to define an induced topological structure and an atlas also on $T^m\mathcal{M}$, whose local coordinate system now will be  represented by $(\boldsymbol{X}_0,...,\boldsymbol{X}_m)$, where $\boldsymbol{X}_0$ refers to $\mathcal{M}$ and each $\boldsymbol{X}_i$ to $T^i\mathcal{M}$, for all $i\in\left\{1,...,m\right\}$. 

In order to employ our framework, we should assure that there exists an energy function equation involving both the physical quantity $\mathbb{E}$ and the local coordinate system $(\boldsymbol{X}_0,...,\boldsymbol{X}_m)$, since this request represents the crucial condition which allows us to simplify considerably the integration process. Furthermore, once we have found a primitive function $V$ of the \qm{lifted} differential one-form $\boldsymbol{\beta}$ with respect to the set of coordinates $\boldsymbol{X}_i$ (with fixed $i\in\left\{0,...,m\right\}$), we could define, in a similar way as in Sec. \ref{sec:ex_form}, the \emph{$i$-th \qm{lifted} vertical exterior derivative} $\boldsymbol{{\rm d^{V,i}}}$ as \cite{Kobayashi1996}
\begin{equation} \label{eq:vertical_derivative_i}
\begin{aligned}
&\boldsymbol{{\rm d^{V,i}}}F= \frac{\partial F}{\partial X_{i,\alpha}} \boldsymbol{{\rm d}}X_{0,\alpha},\\ 
&\forall F\in \mathcal{C}^k(T^m\mathcal{M},\mathfrak{m}),\ k\ge1.
\end{aligned}
\end{equation}
Therefore, we have
\begin{equation}
-\boldsymbol{{\rm d^{V,i}}}V =    \boldsymbol{\beta}. 
\end{equation}
It is worth stressing the fact that the $1$-st \qm{lifted} vertical exterior derivative $\boldsymbol{{\rm d^{V,1}}}$ coincides with the vertical exterior derivative $\boldsymbol{{\rm d^{V}}}$ defined in Sec. \ref{sec:ex_form}.

    
   \item[($vi$)] another remarkable characteristic is that this method is \emph{metric-independent}, i.e., it is not altered by the background geometry, but acts exclusively on the functional form of the equations at stake. This ensures its full validity in general relativity or, in a broader sense, in any metric theory of gravity.       
\end{itemize}

In order to fully appreciate the formalism we have introduced, we note that it shares strong similarities with the approach originally pursued by D'Alembert and Lagrange in order to analyse (free or constrained) mechanical systems. Indeed, its fundamental feature consists on proving that a limited set of essential physical variables (i.e., the Lagrangian coordinates) are sufficient to determine the dynamics. Our method works analogously except that it deals with dissipative systems and in this reveals its breakthrough value. Indeed, Eq. (\ref{eq:constraint}) permits to recognise the energy $\mathbb{E}$ as the only fundamental physical parameter necessary to express analytically the  $V$ primitive and hence describe the dynamics. 


\section{Application of the energy formalism to a simple physical example}
\label{sec:TPE}
In this section we provide a simple physical example in order to show how the \emph{energy formalism} operates on a familiar case. The analysis of the PR effect will be carried out in Sec. \ref{sec:AEF}. 

Let us consider the one-dimensional motion of a test particle of mass $m$ in classical mechanics through a dissipative quadratic medium characterized by the dissipative parameter $\alpha(x)$, viewed as a general smooth function of the $x$ coordinate. The equation of motion reads as  
\begin{equation} \label{eq:EoMC}
m\ddot{x}=-\alpha(x)\dot{x}^2,
\end{equation}
where the dissipative quadratic force is the smooth function
\begin{equation} \label{eq:dissipative_example}
F=-\alpha(x)\dot{x}^2,
\end{equation}
and the dot means time derivative. 

The Rayleigh dissipative potential $V$ can be easily found through standard techniques.  Therefore, we have at once
\begin{equation} \label{eq:EDRF}
F=-\frac{\partial V}{\partial \dot{x}},\quad \Rightarrow\quad V=\alpha(x)\frac{\dot{x}^3}{3}-\alpha(x_0)\frac{\dot{x}_0^3}{3},
\end{equation}
where $x_0$ and $\dot{x}_0$ are the initial position and velocity, respectively.

We can study the same example by applying the \emph{energy formalism}. First of all, the real manifold $\mathcal{M}$ is represented by the line segment described by test particle, which is trivially a simply connected domain (which in turn implies that also $T\mathcal{M}$ is simply connected). In addition, we have a trivial atlas given by intervals of the $x$-axis and the only local coordinate employed is labelled by $x$. Furthermore, the metric is flat and a natural Lebesgue measure is considered. Therefore, the considered example satisfies all the assumptions stated in Sec. \ref{sec:geo_sett}.  

The semi-basic one-form associated to the dissipative force (\ref{eq:dissipative_example}) is given by (cf. Eq. (\ref{differential_oneform_1}))
\begin{equation}
\boldsymbol{\omega}(x,\dot{x})=F(x,\dot{x})\,\boldsymbol{{\rm d}}x\equiv-\alpha(x)\dot{x}^2 \boldsymbol{{\rm d}}x.
\end{equation}
It is not difficult to show that $\boldsymbol{\omega}$ is closed under the vertical exterior derivative (i.e., $\boldsymbol{{\rm d^V}}\boldsymbol{\omega} = 0$) and hence that it is also exact, i.e., Eq. (\ref{eq: differential_of_V}) holds (the formal details can be found in Sec. \ref{sec:ex_form}).

The first crucial requirement of the \emph{energy formalism} consists in singling out the physical quantity dissipated by the system, i.e., the energy function (\ref{eq:constraint}). In this case, such variable is represented by the test particle kinetic energy, which is given by
\begin{equation} \label{eq:KETP}
\mathbb{E}=\frac{1}{2}m\dot{x}^2.
\end{equation} 
The second important step is the derivation of the energy operator (\ref{eq:trader}). By applying the chain rule to the derivatives, we obtain 
\begin{equation} \label{eq:CR}
\frac{\partial }{\partial \dot{x}}=\frac{\partial \mathbb{E}}{\partial \dot{x}}\frac{\partial }{\partial \mathbb{E}}\equiv m\dot{x}\frac{\partial }{\partial \mathbb{E}}, \quad \Rightarrow\quad \frac{\partial }{\partial \dot{x}}=\sqrt{2m\mathbb{E}}\frac{\partial }{\partial \mathbb{E}},
\end{equation}
where we have substituted $\dot{x}=\sqrt{2\mathbb{E}/m}$ through Eq. (\ref{eq:KETP}). 

Finally, the third fundamental step is to write the dissipative force in terms of the dissipated energy $\mathbb{E}$ (cf. Eq. (\ref{eq:constraint_2})). Since in this example Eq. (\ref{eq:KETP}) is easily invertible, we can express $\dot{x}$ in terms of $\mathbb{E}$, as we have already done in Eq. (\ref{eq:CR}). In this way, the dissipative force (\ref{eq:dissipative_example}) can be written as
\begin{equation} \label{eq:FE}
F=-\alpha(x)\frac{2\mathbb{E}}{m}.
\end{equation}
Therefore, the Rayleigh potential $V$ is defined by (cf. Eq. (\ref{eq:primitive}))
\begin{equation} \label{eq:inter}
-\sqrt{2m\mathbb{E}}\frac{\partial V}{\partial \mathbb{E}}=F.
\end{equation}
Multiplying both members of Eq. (\ref{eq:inter}) by the opportune function $B(x,\dot{x})=-1/(m\dot{x})\equiv-1/\sqrt{2m\mathbb{E}}$ (cf. Eq. (\ref{eq:mult})), we obtain (cf. Eq. (\ref{eq:solV}))
\begin{equation} \label{eq:inter2}
\frac{\partial V}{\partial \mathbb{E}}=\frac{F}{-\sqrt{2m\mathbb{E}}}.
\end{equation}
Integrating Eq. (\ref{eq:inter2}) with respect to the energy $\mathbb{E}$, we get (see Eq. (\ref{eq:pot_E}))
\begin{equation} \label{eq:RPF}
V=\frac{2\alpha(x)}{3}\sqrt{\frac{2}{m^3}}\left(\mathbb{E}^{3/2}-\mathbb{E}_0^{3/2}\right)+f(x,\dot{x}),
\end{equation}
where $\mathbb{E}_0=1/2m\dot{x}^2_0$. In this way we have determined the Rayleigh potential in terms of the dissipated energy $\mathbb{E}$ and our ignorance on its dependence on the velocity field has been hidden in the function $f(x,\dot{x})$. It should be stressed that in this trivial case we have easily integrated the force with respect to the energy, but in general this is not a simple task. In order to  determine the function $f(x,\dot{x})$, we first substitute Eq. (\ref{eq:KETP}) in (\ref{eq:RPF}), and then we derive $V$ with respect to the velocity field $\dot{x}$ (see Sec. \ref{sec:prim_en}). In this way, we have
\begin{equation}
-\alpha(x)\dot{x}^2+\frac{\partial f(x,\dot{x})}{\partial \dot{x}}=-\alpha(x)\dot{x}^2,\ \Rightarrow\ \frac{\partial f(x,\dot{x})}{\partial \dot{x}}=0.
\end{equation}
Therefore, our procedure agrees with the result of Eq. (\ref{eq:EDRF}), as it should be expected. 

The example provided in this section clearly shows the fundamental steps underlying the \emph{energy formalism}.
The main feature is represented by the quadratic trend of the energy function $\mathbb{E}$, which permitted to easily invert $\mathbb{E}$ and $\dot{x}$ in Eqs. (\ref{eq:FE}) and (\ref{eq:RPF}), allowing a considerable simplification of the calculations. 
If we analyze the three-dimensional case, the calculations performed through the \emph{energy formalism} are the same. However, such an example allows to appreciate the reduction of the involved variables in the integration process. The one-dimensional model has been chosen  to overcome lengthier calculations and to go straight to the heart of the proposed method.

\section{Application of the energy formalism to the general relativistic PR effect}
\label{sec:AEF}
The motion of test particles, like dust grains or gas clouds \cite{Liou1995,Kimura2002,Klavoka2013}, meteors \cite{Wyatt1950,Yajima2014}, accretion disk matter elements \cite{Rafikov2011,Lancova2017}, around radiating massive sources is strongly affected by gravitational and radiation fields. From X-ray observational data, it is well known that the radiation field exerts an outward-directed (with respect to the radiating source) force against the gravitational pull \cite{Rybicki1986,Frank2002}. However, other perturbing effects must be taken into account, since they might play a fundamental role in altering and driving the test particle trajectories, like radiative heating, magnetic interactions with charged particles, quantum effects, and so forth. In particular, if we consider small not-charged test particles and short time intervals (order of seconds), an important effect to be considered is the PR drag force \cite{Ballantyne2004,Ballantyne2005,Worpel2013,Ji2014,Keek2014,Worpel2015}. 

This phenomenon occurs each time the radiation field invests the test particle, raising up its temperature, which for the Stefan-Boltzmann law starts remitting radiation. This process of absorption and re-emission of radiation generates a recoil force opposite to the test body orbital motion \cite{Burns1979}. Such mechanism removes thus very efficiently angular momentum from the test particle, forcing it to spiral inward or outward depending on the radiation pressure intensity. This force is associated with the action of electromagnetic radiation on a moving spherical body of relatively small size \cite{Klacka2014}. Indeed, a fundamental hypothesis underpinning the PR effect relies on the spherical symmetric distribution of matter inside the test particle. It is assumed that spherical test particles are good approximations of real and arbitrarily shaped bodies \cite{Burns1979,Klacka2014}.

This effect was first studied by J. H. Poynting in 1903 \cite{Poynting1903}, in the context of radiation processes occurring in the Solar System. He formulated the governing equations in the framework of Newtonian gravity and classical physics. However, such developments were not deeply investigated, because at that time the technology providing the needed observational data to validate his intuitions was not available. In 1937, H. P. Robertson extended the classical test body equations of motion to the special relativity frame, considering once again Newtonian gravity model \cite{Robertson1937}. Only seventy years later, in 2009 -- 2011, Bini \emph{et al.} set down previous calculations within a general relativistic pattern by analysing stationary and axially symmetric spacetimes \cite{Bini2009,Bini2011}. After that, PR model has also been extended to three dimensional space in Kerr metric (see Ref. \cite{Defalco20183D}, for details).

Recently, a Lagrangian formulation of the PR effect has been proposed \cite{Defalco2018}. The novel aspects of such approach consists in the introduction of the general relativistic version of Rayleigh dissipation function (from which the PR drag force can be derived straightforwardly) and the use of the integrating factor method (which enlarges the set of exact differential $m$-forms). 

In this section, we exploit the \emph{energy formalism} in order to derive, in a proper way, the general relativistic Rayleigh potential for the PR effect, thus showing the concrete power of this new approach. 
\subsection{Existence of general relativistic Rayleigh potential}
\label{sec:expo}
In this section, we describe some further developments of the framework illustrated in Ref. \cite{Defalco2018}. First of all, we briefly explain the general relativistic PR effect (Sec. \ref{sec:GRPReffect}), afterwards we determine the integrating factor (Sec. \ref{sec:intfact}), and its expression in the classical limit (Sec. \ref{sec:WFA}).

\subsubsection{General relativistic PR effect}
\label{sec:GRPReffect}
The general relativistic PR effect describes the dynamics of a test particle moving with a timelike velocity $\boldsymbol{U}$ around a rotating (or a static) compact object under the influence of a gravitational field, described by the Kerr metric (or the Schwarzschild metric)\footnote{Here, we consider metrics with signature +2, therefore in our notations a timelike vector $v^\alpha$ has norm $v^\alpha v_\alpha=-1$.}in a coordinates system $\boldsymbol{X}$, the radiation pressure, and the radiation drag force. The test particle equations of motion are $a(\boldsymbol{X},\boldsymbol{U})^\alpha=F_{\rm (rad)}(\boldsymbol{X},\boldsymbol{U})^\alpha$, where $a(\boldsymbol{X},\boldsymbol{U})^\alpha$ is the test particle acceleration and $F_{\rm (rad)}(\boldsymbol{X},\boldsymbol{U})^\alpha$ is the radiation force per unit mass, including the radiation pressure and the PR effect. Following the same line of reasoning of J. H. Poynting and H.P.  Robertson \cite{Poynting1903,Robertson1937}, we write the equations of motion first in the test particle rest frame and then in the static observer frame located at infinity. To this aim, we exploit the \emph{relativity of observer splitting formalism}, which represents a powerful method in general relativity to distinguish the gravitational effects from the fictitious forces arising from the relative motion of two non-inertial observers \cite{Jantzen1992,Bini1997a,Bini1997b,Defalco2018}. Such formalism allows us to derive the test particle equations of motion in the reference frame of the static observer located at infinity as a set of coupled first order differential equations \cite{Bini2009,Bini2011,Defalco20183D}. The radiation force is modelled as a pure electromagnetic field, where the photons move along null geodesics on the background spacetime. The stress-energy tensor reads as \cite{Bini2009,Bini2011,Defalco2018,Defalco20183D}
\begin{equation} \label{eq:set}
T^{\alpha\beta}=\Phi^2 k^\alpha k^\beta,
\end{equation}
where $k^\alpha$, which is a function of the local coordinates $\boldsymbol{X}$ only, denotes the photon 4-momentum satisfying the conditions $k_\alpha k^\alpha=0$ and $k^\beta \nabla_\beta k^\alpha=0$, whereas  $\Phi$ represents a parameter related to the radiation field intensity. Therefore, the radiation force $F_{\rm (rad)}(\boldsymbol{X},\boldsymbol{U})^\alpha$ is given by
\begin{equation} \label{eq:radforce1}
\begin{aligned}
F_{\rm (rad)}(\boldsymbol{X},\boldsymbol{U})^\alpha&\equiv-\tilde{\sigma}\mathcal{P}(\boldsymbol{U})^\alpha_{\;\beta} T^\beta_{\;\nu} U^\nu\\
&=-\tilde{\sigma}\Phi^2\left(k^\alpha k_\nu U^\nu+U^\alpha U_\beta k^\beta k_\nu U^\nu\right),
\end{aligned}
\end{equation}
where $\mathcal{P}(\boldsymbol{U})^\alpha_{\; \beta}=\delta^\alpha_{\; \beta} +U^\alpha U_\beta$ is the projection operator on the spatial hypersurface orthogonal to $\boldsymbol{U}$, $\tilde{\sigma}=\sigma/m$ with $\sigma$ the Thomson scattering cross section describing the radiation field-test particle interaction and $m$ the test particle mass. Since the factor $-\tilde{\sigma}\Phi^2$ is a constant with respect to the test particle velocity field $\boldsymbol{U}$, we can ease the notations by considering only
\begin{equation} \label{eq:radforce2}
\tilde{F}_{\rm (rad)}(\boldsymbol{X},\boldsymbol{U})^\alpha\equiv k^\alpha k_\nu U^\nu+U^\alpha U_\beta k^\beta k_\nu U^\nu.
\end{equation}

\subsubsection{Integrating factor}
\label{sec:intfact}
The radiation force $\tilde{F}_{\rm (rad)}(\boldsymbol{X},\boldsymbol{U})^\alpha$ depends non-linearly on the test particle velocity field $\boldsymbol{U}$, therefore we check whether it can be expressed in terms of the Rayleigh potential $V$, i.e., $\tilde{F}_{\rm (rad)}(\boldsymbol{X},\boldsymbol{U})^\alpha=\partial V/\partial U_\alpha$ \cite{Goldstein2002}. It is important to note that the components of (\ref{eq:radforce2}) can be seen as the components of a differential semi-basic one-form $\boldsymbol{\omega}(\boldsymbol{X},\boldsymbol{U})=\tilde{F}_{\rm (rad)}(\boldsymbol{X},\boldsymbol{U})^\alpha \boldsymbol{{\rm d}}X_\alpha$, which is defined over the simply connected domain $T\mathcal{M}$. Indeed, the base spacetime manifold $\mathcal{M}$ is represented by the whole space outside the compact object, including the event horizon, times the time line, whereas all the fibers $T_p\mathcal{M}$ in $p\in\mathcal{M}$ are $n$-dimensional hypercubes, since the limit velocity coincides with the speed of light (see Sec. \ref{sec:ex_form}). Since the cross derivatives of $\tilde{F}_{\rm (rad)}(\boldsymbol{X},\boldsymbol{U})^\alpha$ are not equal, i.e., $\partial \tilde{F}_{\rm (rad)}(\boldsymbol{X},\boldsymbol{U})^\alpha/\partial U_\lambda\neq \partial \tilde{F}_{\rm (rad)}(\boldsymbol{X},\boldsymbol{U})^\lambda/\partial U_\alpha$, the semi-basic one-form turns out to be not exact \cite{Defalco2018}. However, the introduction of an integrating factor $\mu$ could make the differential semi-basic one-form closed in its domain of definition, ensuring thus that $\boldsymbol{\omega}(\boldsymbol{X},\boldsymbol{U})$ is exact. The components of the semi-basic one-form will be now represented by $\mu \tilde{F}_{\rm (rad)}(\boldsymbol{X},\boldsymbol{U})^\alpha$ and the condition according to which this \qm{upgraded differential semi-basic one-form} is closed yields
\begin{equation} \label{eq:difeqformu}
\begin{aligned}
0&=\left(-k^\alpha\frac{\partial\mu}{\partial U_\lambda}+k^\lambda\frac{\partial\mu}{\partial U_\alpha}\right)\\
&+U^\alpha\left(\frac{\partial \mu}{\partial U_\lambda}k^\beta U_\beta+2\mu k^\lambda\right)\\
&-U^\lambda\left(\frac{\partial \mu}{\partial U_\alpha}k^\beta U_\beta+2\mu k^\alpha\right).\\
\end{aligned}
\end{equation}
This in turn implies that $\mu$ should solve simultaneously the following two differential equations:
\begin{eqnarray} 
&&-k^\alpha\frac{\partial\mu}{\partial U_\lambda}+k^\lambda\frac{\partial\mu}{\partial U_\alpha}=0, \label{eq:difeqformu1}\\
&&\frac{\partial \mu}{\partial U_\lambda}k^\beta U_\beta+2\mu k_\lambda=0. \label{eq:difeqformu2}
\end{eqnarray}
The radiation force (\ref{eq:radforce2}) can be split into two parts
\begin{equation} \label{eq:splitforce}
\tilde{F}_{\rm (rad)}(\boldsymbol{X},\boldsymbol{U})^\alpha=\mathbb{F}_{\rm C}(\boldsymbol{X},\boldsymbol{U})^\alpha+\mathbb{F}_{\rm NC}(\boldsymbol{X},\boldsymbol{U})^\alpha,
\end{equation}
where
\begin{eqnarray}
\mathbb{F}_{\rm C}(\boldsymbol{X},\boldsymbol{U})^\alpha&\equiv& T^\alpha_{\;\nu} U^\nu=-k^\alpha \mathbb{E}(\boldsymbol{X},\boldsymbol{U}), \label{eq:twoforces1}\\
&&\notag\\
\mathbb{F}_{\rm NC}(\boldsymbol{X},\boldsymbol{U})^\alpha&\equiv& U^\alpha U_\beta T^\beta_{\;\nu} U^\nu=\mathbb{E}(\boldsymbol{X},\boldsymbol{U})^2 U^\alpha,\label{eq:twoforces2}
\end{eqnarray}
with 
\begin{equation} \label{particle_energy_1}
\mathbb{E}(\boldsymbol{X},\boldsymbol{U})\equiv \mathbb{E}=-k_\beta U^\beta, 
\end{equation}
representing the test particle energy \cite{Bini2009,Bini2011,Defalco2018} (more details regarding $\mathbb{E}$ will be given in Sec. \ref{sec:LOGPOT}). We refer to $\mathbb{F}_{\rm C}(\boldsymbol{X},\boldsymbol{U})^\alpha$ as the ``conservative'' part of the radiation force, due to its property to admit a primitive function without considering an integrating factor, since it depends linearly on $U^\alpha$; whereas $\mathbb{F}_{\rm NC}(\boldsymbol{X},\boldsymbol{U})^\alpha$ stands for the ``non-conservative'' components of the radiation force, because the related primitive can be determined only through the introduction of the integrating factor $\mu$. 

At this stage, we would like to determine a common integrating factor for both the components of the radiation force. However, it is noteworthy to stress that this request is not trivial at all. In fact, in principle we might come up with two different integrating factors $\mu_1$ and $\mu_2$, where $\mu_1$ is related to $\mathbb{F}_{\rm C}(\boldsymbol{X},\boldsymbol{U})^\alpha$ and solves Eq. (\ref{eq:difeqformu1}), while $\mu_2$ is associated with $\mathbb{F}_{\rm NC}(\boldsymbol{X},\boldsymbol{U})^\alpha$ and solves Eq. (\ref{eq:difeqformu2}). Therefore, the system of differential equations (\ref{eq:difeqformu1}) and (\ref{eq:difeqformu2}) admits in general two distinct solutions\footnote{In the most general case it may even happen that some of the differential equations, defining the various integrating factors, might not admit any solution at all.} ($\mu_1\neq\mu_2$). For instance, for the conservative components $\mathbb{F}_{\rm C}(\boldsymbol{X},\boldsymbol{U})^\alpha$ the function $\mu=\mbox{const}$ clearly represents a solution of (\ref{eq:difeqformu1}). Therefore, the possibility to find a unique solution, different from the trivial one $\mu=\mbox{const}$, for both $\mathbb{F}_{\rm NC}(\boldsymbol{X},\boldsymbol{U})^\alpha$ and $\mathbb{F}_{\rm C}(\boldsymbol{X},\boldsymbol{U})^\alpha$ is not so obvious \emph{a priori}. However, as it can be easily checked from Eqs. (\ref{eq:difeqformu1}) and (\ref{eq:difeqformu2}), the PR effect exhibits the peculiar propriety to have one common integrating factor for the two components (i.e., $\mu_1=\mu_2 \equiv \mu$), which, up to a constant, reads as\footnote{In Eq. (\ref{eq:if}) we have corrected a little error occurred in Ref. \cite{Defalco2018}.}
\begin{equation} \label{eq:if}
\mu =\frac{1}{\mathbb{E}^2}.
\end{equation}
The use of an integrator factor permits to guarantee, in a non-intuitive manner, existence and uniqueness (up to a constant term) of the Rayleigh potential. In addition, this represent a powerful method in metric theories of gravity, where the coupling between external dissipative forces and geometrical background gives rise to non-linear functions.

\subsubsection{Classical limit of the integrating factor}
\label{sec:WFA}
As we have just pointed out, the integrating factor (\ref{eq:if}) is defined up to a constant (with respect to the velocity field $\boldsymbol{U}$) which can be determined in the \emph{classical limit} (weak field approximation, $M/r\rightarrow0$, and non-relativistic velocities, $\nu/c\rightarrow0$, \cite{Defalco2018}). By employing the Schwarzschild metric in the equatorial plane $\theta=\pi/2$,
\begin{equation} \label{Schwarzschild_metric_1}
g_{\mu \nu}=\rm{diag}\left[-\left(1-\frac{2M}{r}\right),\left(1-\frac{2M}{r}\right)^{-1},r^2,r^2\right],
\end{equation}
the test particle and the photon velocities read, respectively, as \cite{Bini2009,Defalco2018,Chandrasekhar1992}\footnote{Due to the spherical symmetry of the metric, we are allowed, without loss of generality, to reduce the problem to a two dimensional setting so that all calculations are easily performed.}
\begin{eqnarray} 
U^\alpha&=&\left[\frac{\gamma}{\sqrt{1-\frac{2M}{r}}},\frac{\gamma\nu\sin\alpha}{\left(\sqrt{1-\frac{2M}{r}}\right)^{-1}},0,\frac{\gamma\nu\cos\alpha}{r}\right], \label{eq:vel-phot1}\\
k_\alpha&=&E_{\rm p}\left[-1,\frac{1}{1-\frac{2M}{r}},0,0\right], \label{eq:vel-phot2}
\end{eqnarray}
where $\gamma$ is the Lorentz factor, $\alpha$ and $\nu$ represent, respectively, the azimuthal angle and the module of the test particle velocity in the spatial hypersurface orthogonal to the zero angular momentum observers (ZAMOs), $E_{\rm p}=-k_t$ is the photon energy. 
Note that, without loss of generality, we have considered a radial radiation photon impact parameter (see \cite{Bini2009,Bini2011,Defalco2018}, for further details). 

Bearing in mind Eqs. (\ref{particle_energy_1}), (\ref{Schwarzschild_metric_1}), (\ref{eq:vel-phot1}), and (\ref{eq:vel-phot2}), it is easy to show that in the classical limit
\begin{equation} \label{eq:Elimit}
\mathbb{E}\approx E_{\rm p}\left(1-\dot{r}\right),
\end{equation}
where we have decided to neglect, from now on, all terms containing the factor $M/r$, since classically they give a higher-order contribution to the radiation force. We note that the classical Rayleigh potential can be easily recovered without the introduction of an integrating factor (see Ref. \cite{Defalco2018}), hence in this limit $\mu=1$. Since $\mu\sim\mbox{const}/(\mathbb{E}^2)$, this leads immediately to choose the constant term equals to $E_{\rm p}^2$, so that Eq. (\ref{eq:if}) can now be recasted as
\begin{equation} \label{eq:if2}
\mu =\frac{E_{\rm p}^2}{\mathbb{E}^2}.
\end{equation}
The appearance of a constant term having the physical dimension of the square of an energy could also be expected on general grounds, since we require a dimensionless integrating factor. 

\subsection{General relativistic Rayleigh potential}
\label{sec:GRRPO_sol}
In this section we exploit the \emph{energy formalism} to determine the general relativistic Rayleigh potential of the PR effect. We apply extensively the description enlightened in Sec. \ref{sec:EF} to a concrete astrophysical model in order to show the strength of this new approach. 

\subsubsection{Preliminary}
\label{sec:preliminary}
The geometrical setup of the PR effect fully respects the framework delineated in Sec. \ref{sec:geo_sett}. Indeed, the manifold $\mathcal{M}$ is represented by the three-dimensional space outside the compact object, included the event horizon, times the time line. In other words, $\mathcal{M}$ coincides with Schwarzschild or Kerr spacetimes, which can always be split in space and time. Moreover, such spacetimes benefit of all the differential and topological proprieties required in Sec. \ref{sec:geo_sett}. Therefore, $\mathcal{M}$ is a four-dimensional, real, topological Hausdorff, differential, pseudo Riemannian (or Lorentzian), simply connected, and asymptotically flat manifold \cite{Misner1973}. Furthermore, the standard Lebesgue measure gets multiplied by a metric factor $\sqrt{-g}$, where $g$ is the determinant of the Schwarzschild or Kerr metric. The local coordinates system $\boldsymbol{X}$ coincides with the Boyer-Lindquist (for Kerr metric) or spherical (for Schwarzschild metric) coordinates. Since we would like to investigate the Rayleigh potential, our coordinates system for $T\mathcal{M}$ will be represented by $\boldsymbol{X}$ and the test particle velocity field $\boldsymbol{U}$ (see Sec. \ref{sec:geo_sett}). 

The components of the differential semi-basic one-form (\ref{differential_oneform_1}) are now expressed in terms of the components of the radiation force (\ref{eq:radforce2}). In addition, $\boldsymbol{\omega}(\boldsymbol{X},\boldsymbol{U})$ satisfies all the regularity conditions required by the \emph{energy formalism}. In particular, as pointed in Sec. \ref{sec:expo}, the introduction of the integrating factor (\ref{eq:if}) (or equivalently (\ref{eq:if2})) makes $\boldsymbol{\omega}$ closed and hence, according to our hypotheses regarding the topological properties of its domain $T\mathcal{M}$, exact (see Sec. \ref{sec:geo_sett}). Therefore, it makes sense the research of a potential function $V(\boldsymbol{X},\boldsymbol{U})$ (i.e., the Rayleigh potential) such that, in analogy with what has been done in Eq. (\ref{eq: differential_of_V}), we have
\begin{equation}
-\boldsymbol{{\rm d^V}}V(\boldsymbol{X},\boldsymbol{U})=\mu \, \boldsymbol{\omega}(\boldsymbol{X},\boldsymbol{U}).
\end{equation}
We split the potential function $V(\boldsymbol{X},\boldsymbol{U})$ in two parts according to (see Eqs. (\ref{eq:splitforce}), (\ref{eq:twoforces1}), and (\ref{eq:twoforces2}))
\begin{equation} \label{eq:teopot}
\tilde{F}_{\rm (rad)}(\boldsymbol{X},\boldsymbol{U})^\alpha=-\frac{1}{\mu}\frac{\partial V}{\partial U_\alpha}=-\frac{1}{\mu}\frac{\partial (\mathbb{V}_{\rm C}+\mathbb{V}_{\rm NC})}{\partial U_\alpha},
\end{equation}
where
\begin{eqnarray} 
\mu\mathbb{F}_{\rm C}(\boldsymbol{X},\boldsymbol{U})^\alpha&=&-\frac{\partial \mathbb{V}_{\rm C}}{\partial U_\alpha},\label{eq:teopot21}\\
&&\notag\\
\mu\mathbb{F}_{\rm NC}(\boldsymbol{X},\boldsymbol{U})^\alpha&=&-\frac{\partial \mathbb{V}_{\rm NC}}{\partial U_\alpha}. \label{eq:teopot22}
\end{eqnarray}
We will determine $\mathbb{V}_{\rm C}$ and $\mathbb{V}_{\rm NC}$ in Secs. \ref{sec:PC} and \ref{sec:PNC}, respectively.

For the PR effect, the \emph{energy function} (\ref{eq:constraint}) is given by Eq. (\ref{particle_energy_1}). In this case, the \emph{energy hypersurface} (Sec. \ref{sec:en_coinst}) is represented by a six-dimensional hypersurface embedded in the eight-dimensional space $T\mathcal{M}$ (in the case of the three-dimensional PR effect \cite{Defalco20183D}) or four-dimensional in the six-dimensional space $T\mathcal{M}$ (for the two-dimensional PR effect \cite{Bini2009,Bini2011}). This is due to the fact that besides Eq. (\ref{particle_energy_1}), we can exploit the additional \emph{module constraint}, i.e., $U_\alpha U^\alpha=-1$. The transformation rule (\ref{eq:trader}) of the derivative operator in terms of the energy reads as
\begin{equation} \label{eq:der}
\frac{\partial}{\partial U_\alpha}=\frac{\partial \mathbb{E}}{\partial U_\alpha} \frac{\partial}{\partial \mathbb{E}}=-k^\alpha \frac{\partial}{\partial \mathbb{E}}.
\end{equation}  
The components of the differential semi-basic one-form $\boldsymbol{\omega}$ can be expressed in terms of the energy (cf. Eq. (\ref{eq:constraint_2})) and are given by Eqs. (\ref{eq:twoforces1}) and (\ref{eq:twoforces2}).

\subsubsection{Conservative potential}
\label{sec:PC}
In accordance with our definitions (see Eqs. (\ref{eq:primitive}), (\ref{eq:twoforces1}), (\ref{eq:if}), and (\ref{eq:teopot21})), the potential $\mathbb{V}_{\rm C}$ is defined by
\begin{equation} \label{eq:potC}
-\frac{\partial \mathbb{V}_{\rm C}}{\partial U_\alpha}=-\frac{k^\alpha}{\mathbb{E}}\quad
\Leftrightarrow\quad
-\frac{\partial \mathbb{V}_{\rm C}}{\partial \mathbb{E}}=\frac{1}{\mathbb{E}}.
\end{equation}
In this case, once the energy operator (\ref{eq:der}) has been exploited, the function $k^\alpha$ simplifies on both members of the second equation in (\ref{eq:potC}). Therefore, the opportune function $B(\boldsymbol{X},\boldsymbol{U},\mathbb{E})_\alpha$ occurring in Eq. (\ref{eq:mult}) becomes trivial, since it can be given by any non-vanishing function of the local coordinates ($\boldsymbol{X},\boldsymbol{U}$), because it eventually simplifies on both members of the second of (\ref{eq:potC}). Therefore, Eqs. (\ref{eq:mult}) and (\ref{eq:solV}) reduce to the second equation in (\ref{eq:potC}), which, bearing in mind Eq. (\ref{eq:pot_E}), can be easily integrable in terms of the energy $\mathbb{E}$, yielding
\begin{equation} \label{eq:potC2}
\mathbb{V}_{\rm C}=-\ln(\mathbb{E})+f(\boldsymbol{X},\boldsymbol{U}).
\end{equation}
To determine $f(\boldsymbol{X},\boldsymbol{U})$, we need to employ the iterative process outlined in Sec. \ref{sec:prim_en}. Therefore, we have
\begin{equation}
-\frac{\partial \mathbb{V}_{\rm C}}{\partial U_\alpha}=-\frac{k^\alpha}{\mathbb{E}}-\frac{\partial f(\boldsymbol{X},\boldsymbol{U})}{\partial U_\alpha}, 
\end{equation}
which is exactly equal to the corresponding component $\mu \mathbb{F}_{\rm C}(\boldsymbol{X},\boldsymbol{U})^\alpha$ of the radiation force if ${\partial f(\boldsymbol{X},\boldsymbol{U})}/{\partial U_\alpha}=0$. 
We have $f(\boldsymbol{X},\boldsymbol{U})=\mbox{const}$, where the constant will be determined in Sec. \ref{sec:WFAP}. 

\subsubsection{Non-conservative potential}
\label{sec:PNC}
In this case the joint application of Eqs. (\ref{eq:primitive}), (\ref{eq:twoforces2}) and (\ref{eq:teopot22}) gives for potential $\mathbb{V}_{\rm NC}$
 \begin{equation} \label{eq:newPNC}
-\frac{\partial \mathbb{V}_{\rm NC}}{\partial U_\alpha}=U^\alpha\quad
\Leftrightarrow\quad
k^\alpha \, \frac{\partial \mathbb{V}_{\rm NC}}{\partial \mathbb{E}}=U^\alpha.
\end{equation}
In view of Eq. (\ref{eq:mult}), the function $B(\boldsymbol{X},\boldsymbol{U},\mathbb{E})_\alpha$ reduces simply to $U_\alpha$, since the condition $U^\alpha U_\alpha=-1$ jointly with Eq. (\ref{particle_energy_1}) allows us to rearrange the last equation in
\begin{equation} \label{eq:newPNC_2}
\frac{\partial \mathbb{V}_{\rm NC}}{\partial \mathbb{E}} =  \dfrac{1}{\mathbb{E}},
\end{equation}
which is the same as Eq. (\ref{eq:solV}).

Therefore, in formal analogy with Eq. (\ref{eq:pot_E}), integrating Eq. (\ref{eq:newPNC_2}) with respect to $\mathbb{E}$ leads to
\begin{equation} \label{eq:potPNC2}
\mathbb{V}_{\rm NC}=\ln(\mathbb{E})+f(\boldsymbol{X},\boldsymbol{U}).
\end{equation}
Integrating Eq. (\ref{eq:potPNC2}) we determine $f(\boldsymbol{X},\boldsymbol{U})$ as
\begin{equation}
-\frac{\partial \mathbb{V}_{\rm NC}}{\partial U_\alpha}=\frac{k^\alpha}{\mathbb{E}}-\frac{\partial f(\boldsymbol{X},\boldsymbol{U})}{\partial U_\alpha}.
\end{equation}
Comparing the above derivatives with the corresponding components $\mu \mathbb{F}_{\rm NC}(\boldsymbol{X},\boldsymbol{U})^\alpha$ of the radiation force, we find 
\begin{equation} \label{eq:potPNC3}
\begin{aligned}
f(\boldsymbol{X},\boldsymbol{U})&=\int\left(-U^\alpha+\frac{k^\alpha}{\mathbb{E}}\right)dU_\alpha\\
&=-\int U^\alpha\ dU_\alpha-\ln(\mathbb{E}), \;\;\;\; ({\rm no \; sum \; over}\;  \alpha).
\end{aligned}
\end{equation}
Substituting the last expression in Eq. (\ref{eq:potPNC2}), we obtain $\mathbb{V}_{\rm NC}=-\int U^\alpha\ dU_\alpha$. After some algebra following the procedure reported in Sec. \ref{sec:prim_en}, we obtain 
\begin{equation} \label{eq:potPNC4}
\mathbb{V}_{\rm NC}=-\frac{1}{2}U^\alpha U_\alpha +\mbox{const},
\end{equation}
where the integration constant will be determined in Sec. \ref{sec:WFAP}.

In conclusion, in the last sections we have realized how the \emph{energy formalism} has allowed us to compute in a straightforward way the Rayleigh potential of the PR effect, which reads as
\begin{equation} \label{eq:Rayleigh_potential_final}
V\equiv\mathbb{V}_{\rm C}+\mathbb{V}_{\rm NC} =-\left[ \ln(\mathbb{E})+\frac{1}{2}U_\alpha U^\alpha\right]+\mbox{const}.
\end{equation}

\subsubsection{Classical limit of the potential}
\label{sec:WFAP}
At this stage, it is interesting to check whether the Rayleigh potential (\ref{eq:Rayleigh_potential_final}) is consistent with the classical equations first introduced by Poynting and Robertson \cite{Poynting1903,Robertson1937}. 

In the classical limit, the components of the radiation force (\ref{eq:radforce1}) become
\begin{eqnarray}
F_{\rm (rad)}(\boldsymbol{X},\boldsymbol{U})^r&\approx&-\frac{A}{r^2}\left(2\dot{r}-1\right), \label{eq:limF1}\\
F_{\rm (rad)}(\boldsymbol{X},\boldsymbol{U})^\varphi&\approx&-\frac{A}{r^2}\left(r\dot{\varphi}\right), \label{eq:limF2}\\
F_{\rm (rad)}(\boldsymbol{X},\boldsymbol{U})^t&=&\dot{r}F_{\rm (rad)}(\boldsymbol{X},\boldsymbol{U})^r+r\dot{\varphi}F_{\rm (rad)}(\boldsymbol{X},\boldsymbol{U})^\varphi
\notag\\
&\approx&-\frac{A}{r^2}\left(-\dot{r}+2\dot{r}^2+r^2\dot{\varphi}^2\right). \label{eq:limF3}
\end{eqnarray}
We remind that factor $\Phi^2$ occurring in Eq. (\ref{eq:radforce1}) can be written as $\Phi^2=\Phi_0^2/r^2$, where $\Phi_0$ is a constant related to the intensity of the radiation field at the emitting surface, and $A=\tilde{\sigma}\Phi^2_0E^2_{\rm p}$ is the luminosity parameter ranging in the interval $[0,1]$, see Refs. \cite{Bini2009,Bini2011,Defalco20183D}. Thus, we can relate the radiation force with the derivative of the Rayleigh potential (\ref{eq:Rayleigh_potential_final}) and the integrating factor (\ref{eq:if2}) through (cf. Eq. (\ref{eq:radforce2})) 
\begin{equation} \label{eq:startpoint}
\begin{aligned}
F_{\rm (rad)}(\boldsymbol{X},\boldsymbol{U})^\alpha&=
-\frac{\tilde{\sigma}\Phi_0^2}{r^2}\tilde{F}_{\rm (rad)}(\boldsymbol{X},\boldsymbol{U})^\alpha\\
&=\frac{A}{r^2}\frac{\mathbb{E}^2}{E^2_{\rm p}}\frac{\partial V}{\partial U_\alpha},
\end{aligned}
\end{equation}
where the last equality can be obtained after having multiplicated and divided $\tilde{F}_{\rm (rad)}(\boldsymbol{X},\boldsymbol{U})^\alpha$ by $\mu$. The conservative and non-conservative components occurring in the general relativistic Rayleigh potential (\ref{eq:Rayleigh_potential_final}) assume in the classical limit the following form, respectively:
\begin{eqnarray} \label{eq:POTCL1}
\ln(\mathbb{E})&\approx&\ln(E_{\rm p})-\dot{r}-\frac{\dot{r}^2}{2},\label{eq:POTCL1a}\\
\frac{1}{2}U_\alpha U^\alpha&\approx&\frac{1}{2}\left(-1+\dot{r}^2+r^2\dot{\varphi}^2\right). \label{eq:POTCL1b}
\end{eqnarray}
Therefore, in the classical limit we have 
\begin{equation} \label{eq:POTCL2}
V\approx\dot{r}-\frac{1}{2}r^2\dot{\varphi}^2+\left[\frac{1}{2}-\ln (E_{\rm p})\right]+\mbox{const},
\end{equation}
where we can choose 
\begin{equation} \label{eq:constant_term_potential}
\mbox{const}=-\left[\frac{1}{2}-\ln (E_{\rm p})\right],
\end{equation}
to cancel out the term appearing in the square bracket of (\ref{eq:POTCL2}). From Eqs. (\ref{eq:startpoint}) and (\ref{eq:POTCL2}), we have
\begin{equation} \label{eq:startpoint2}
\begin{aligned}
F_{\rm (rad)}(\boldsymbol{X},\boldsymbol{U})^\alpha&\approx\frac{A}{r^2}\left(1-\dot{r}\right)^2\frac{\partial }{\partial U_\alpha}\left(\dot{r}-\frac{1}{2}r^2\dot{\varphi}^2\right),
\end{aligned}
\end{equation}
where Eq. (\ref{eq:Elimit}) has been exploited. Therefore, it is simple to check that Eq. (\ref{eq:startpoint2}) leads immediately to the classical expressions (\ref{eq:limF1}) and (\ref{eq:limF2}) once the underlying derivatives are computed.
Since the components $F_{\rm (rad)}(\boldsymbol{X},\boldsymbol{U})^r$ and $F_{\rm (rad)}(\boldsymbol{X},\boldsymbol{U})^\varphi$ deduced from (\ref{eq:startpoint2}) match correctly with their own classical limit, it is obvious that also Eq. (\ref{eq:limF3}) is straightforwardly satisfied. 

\subsection{Discussion and interpretation of the results}
\label{sec:int_res}
It is extremely important to analyse the results found in the previous sections to focus the attention on several interesting points. Firstly, bearing in mind the factor $-\tilde{\sigma}\Phi^2$ mentioned in Sec. \ref{sec:GRPReffect} and Eqs. (\ref{eq:Rayleigh_potential_final}) and (\ref{eq:constant_term_potential}), the complete analytic expression of the general relativistic Rayleigh potential for the PR effect reads as
\begin{equation} \label{eq:Rayleigh_potential_PR}
V=\tilde{\sigma}\Phi^2\left[\ln\left(\frac{\mathbb{E}}{E_{\rm p}}\right)+\frac{1}{2}\left(U_\alpha U^\alpha+1\right)\right].
\end{equation}
According to Eq. (\ref{eq:dissipation}) and information reported beneath, the classical PR effect amounts to be a \emph{viscous force}, since it is characterized by a linear dependence on the test particle velocity \cite{Poynting1903,Robertson1937}. Rayleigh potential represents physically the energy associated to the system under investigation.
From Ref. \cite{Bini2011}, we have 
\begin{equation} \label{eq:phi_explicit}
\Phi^2=\Phi^2_0\Bigg/ r^2\left[\frac{\rho}{r^3}\left(1-\frac{2aMb}{\rho}\right)^2-
\left(\frac{b\Delta}{\rho}\right)^2\right]^{1/2},
\end{equation}
where $\Delta=r^2-2Mr+a^2$ and $\rho=r^3+a^2r+2a^2M$, $M$ and $a$ being mass and spin of the black hole (BH), respectively. In addition, remembering that $A=\tilde{\sigma}\Phi^2_0E^2_{\rm p}$, and moving
$1/E^2_{\rm p}$ outside of the potential, as in Eq. (\ref{eq:startpoint}), we obtain
\begin{equation} \label{eq:potential_explicit}
V=\frac{A}{r^2}\ \frac{\left[\ln\left(\frac{\mathbb{E}}{E_{\rm p}}\right)+\frac{1}{2}\left(U_\alpha U^\alpha+1\right)\right]}{\left[\frac{\rho}{r^3}\left(1-\frac{2aMb}{\rho}\right)^2-\left(\frac{b\Delta}{\rho}\right)^2\right]^{1/2}}.
\end{equation}

After the publication of Poynting's paper in 1903, a fierce controversy arose in the scientific community,  
which has been originated by the inconsistency of the eponymous effect both with the principles of relativistic mechanics and Maxwell theory of electromagnetism. First attempts to solve such issue were made by J. Larmor and L. Page through aether theory \cite{Larmor1917,Page1918}. The question was partially clarified by H. P. Robertson, who reformulated Poynting's model in the context of special relativity (see Ref. \cite{Robertson1937}, and references therein). Recently, a debate regarding the physical foundations of PR effect between Kla{\v c}ka \emph{et al.} and Burns \emph{et al.} has appeared in the literature \cite{Burns1979,Klacka2014,Burns2014}. 

Einstein theory solves elegantly all delicate difficulties underlying PR model \cite{Bini2009,Bini2011}. In particular, the test particle equations of motion are such that both radiation pressure and PR drag force contributions are included in a single function, i.e., the relativistic radiation force \cite{Klacka2014}. In other words, the general covariance principle prevents any kind of separation between these two terms. On the contrary, such a splitting is admissible only at classical level. The Rayleigh potential reported in Eq. (\ref{eq:potential_explicit}) tremendously support this argument, since it confirms the fact that in a relativistic framework we are not able to distinguish the radiation pressure from the PR effect potential (contrarily to the classical case \cite{Defalco2018}).

Another important feature of the PR model is the dependence of the radiation force on the test particle velocity $\boldsymbol{U}$, see Eq. (\ref{eq:radforce1}). In Sec. \ref{sec:WFAP}, we note that in the classical limit the time component $U^t$ is connected with the radiation pressure, while the spatial components, $U^r,U^\theta,U^\varphi$, are linked to the PR drag force. Such a remark implies that the radiation pressure always enters the dynamics, since $U^t$ never vanishes, while the PR drag force could be turned off whenever the test particle is at rest (see Ref. \cite{Defalco20183D}, and figures therein for details).

\subsubsection{Rayleigh potential and test particle trajectory}
\label{sec:RP_TPT}
\begin{figure*}[th!]
\centering
\includegraphics[scale=0.64]{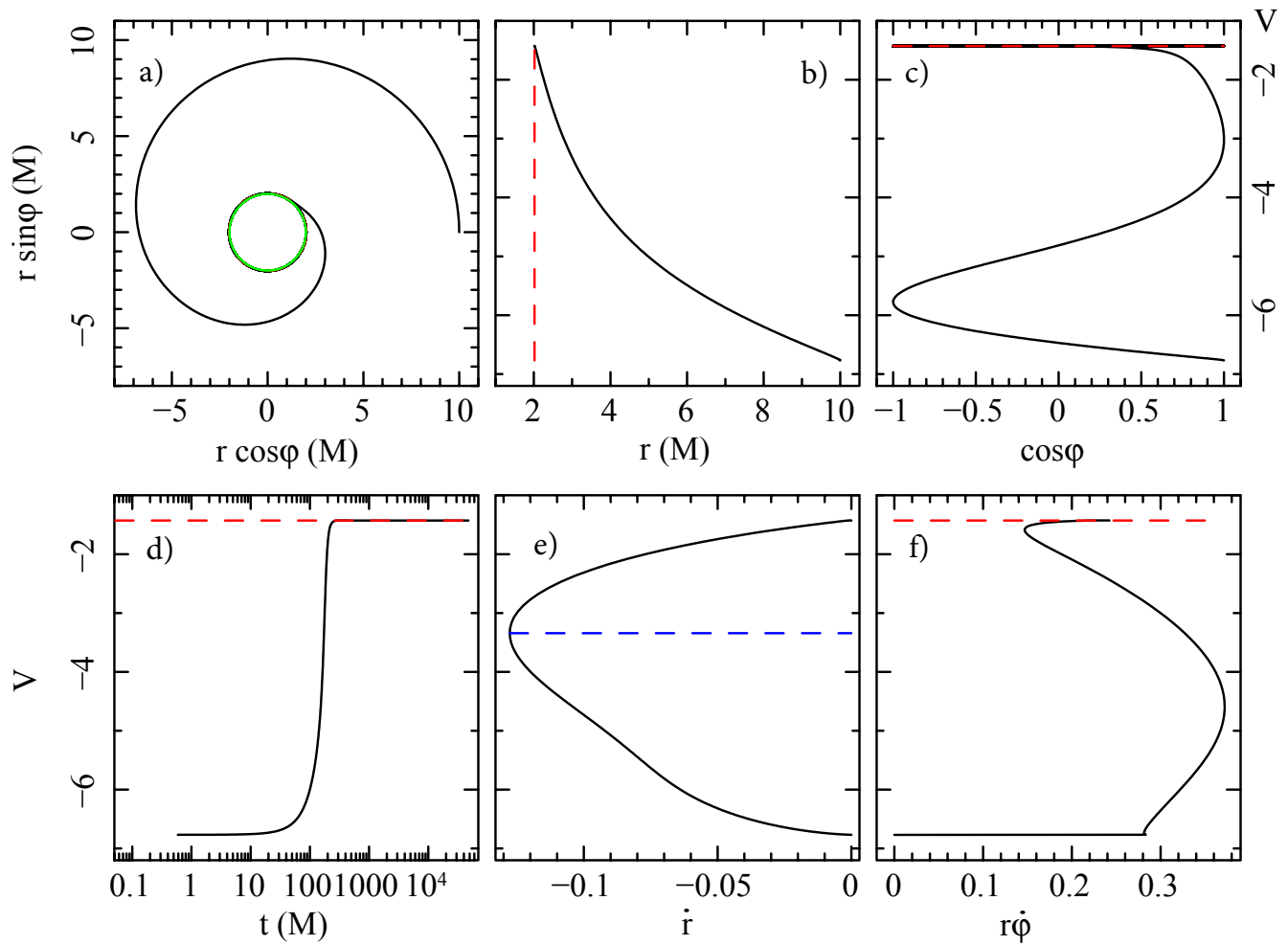}
\caption{Test particle trajectory with the related general relativistic Rayleigh potential (\ref{eq:potential_explicit}) for mass $M=1$ and spin $a=0.1$ of the BH, luminosity parameter $A=0.1$ and photon impact parameter $b=1$. The test particle moves in the spatial equatorial plane with initial position $(r_0,\varphi_0)=(10M,0)$ and velocity $(\nu_0,\alpha_0)=(\sqrt{1/10M},0)$. a) Test particle trajectory spiralling towards the BH and stopping on the critical radius (red dashed line) $r_{\rm (crit)}=2.02M$. The continuous green line is the event horizon radius $r^+_{\rm (EH)}=1.99M$. Rayleigh potential versus b) radial coordinate, c) azimuthal coordinate, d) time coordinate, e) radial velocity, and f) azimuthal velocity. The blue dashed line in panel e) marks the minimum value attained by the radial velocity, corresponding to $\dot{r}=-0.13$. Panels b)--f) must be read from bottom up.}
\label{fig:Fig1}
\end{figure*}

\begin{figure*}[th!]
\centering
\includegraphics[scale=0.64]{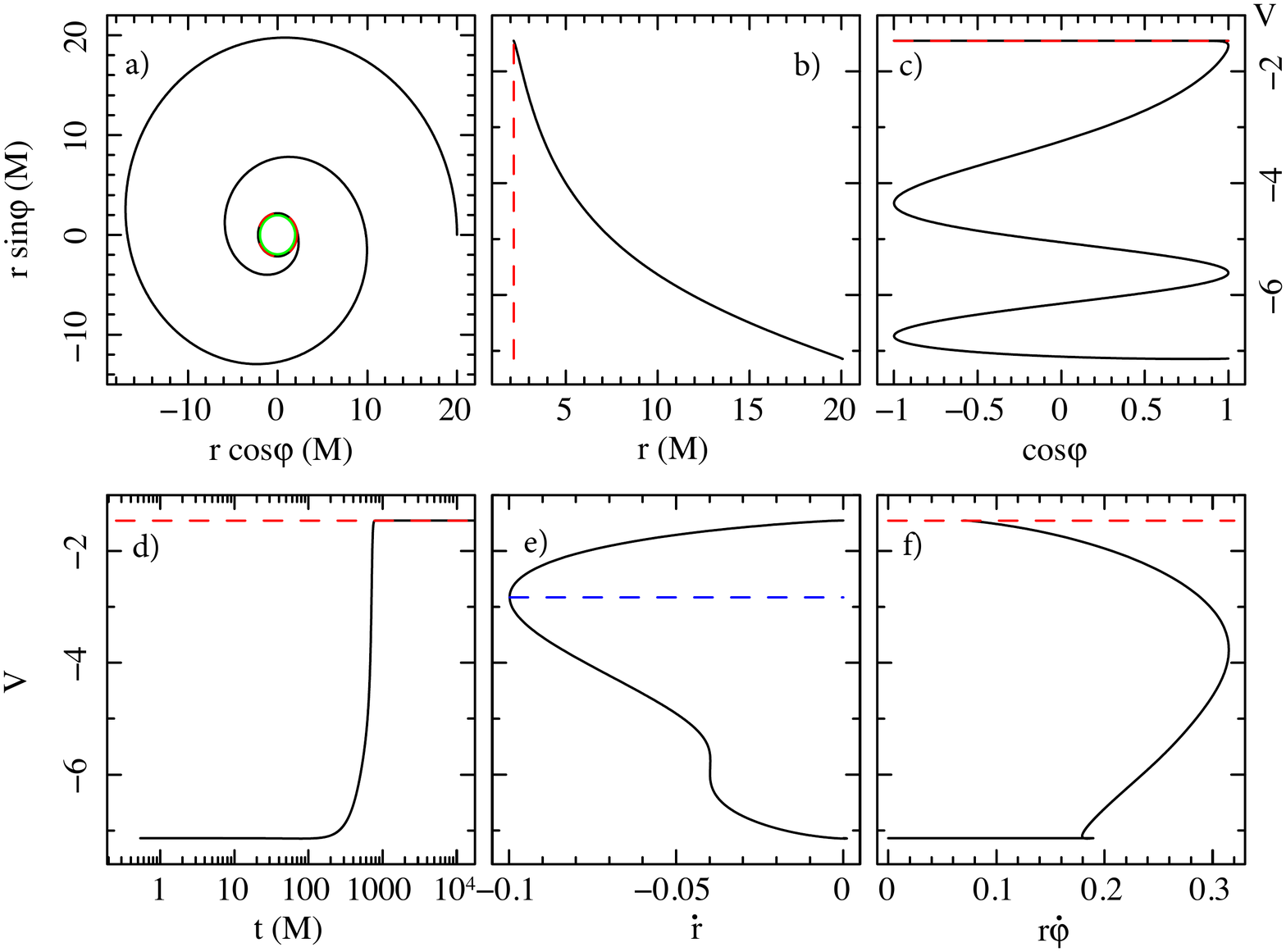}
\caption{Test particle trajectory with the related general relativistic Rayleigh potential (\ref{eq:potential_explicit}) for mass $M=1$ and spin $a=0$ of the BH, luminosity parameter $A=0.3$ and photon impact parameter $b=2$. The test particle moves in the spatial equatorial plane with initial position $(r_0,\varphi_0)=(20M,0)$ and velocity $(\nu_0,\alpha_0)=(0.2,0)$. a) Test particle trajectory spiralling towards the BH and stopping on the critical radius (red dashed line) $r_{\rm (crit)}=2.16M$. The continuous green line is the event horizon radius $r^+_{\rm (EH)}=1.95M$. Rayleigh potential versus b) radial coordinate, c) azimuthal coordinate, d) time coordinate, e) radial velocity, and f) azimuthal velocity. The blue dashed line in panel e) marks the minimum value attained by the radial velocity, corresponding to $\dot{r}=-0.1$. Panels b)--f) must be read from bottom up.} 
\label{fig:Fig2}
\end{figure*}

\begin{figure*}[th!]
\centering
\includegraphics[scale=0.64]{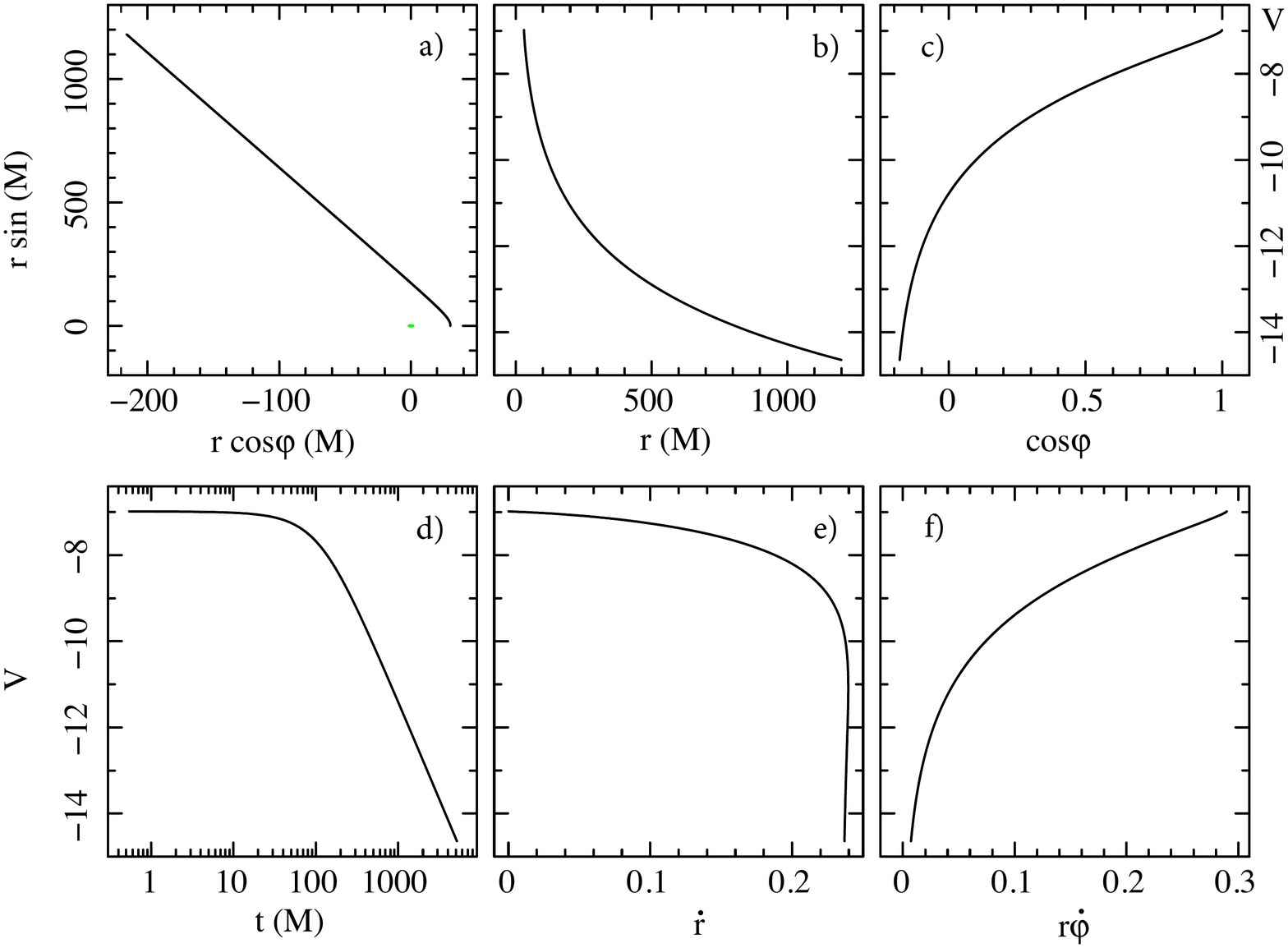}
\caption{Test particle trajectory with the related general relativistic Rayleigh potential (\ref{eq:potential_explicit}) for mass $M=1$ and spin $a=0.8$ of the BH, luminosity parameter $A=0.8$ and photon impact parameter $b=5$. The test particle moves in the spatial equatorial plane with initial position $(r_0,\varphi_0)=(30M,0)$ and velocity $(\nu_0,\alpha_0)=(0.3,0)$. a) Test particle trajectory departing from the BH and approaching spatial infinity. The continuous green line is the event horizon radius $r^+_{\rm (EH)}=1.6M$. Rayleigh potential versus b) radial coordinate, c) azimuthal coordinate, d) time coordinate, e) radial velocity, and f) azimuthal velocity. Panels b)--f) must be read from top down.} 
\label{fig:Fig3}
\end{figure*}

The analytical form of the Rayleigh potential is relevant for its strong correspondence with the test particle trajectory, creating thus a direct link with observations. From Fig. \ref{fig:Fig1}, it is possible to note explicitly this connection. Indeed in panel a), we see that the test particle spirals inward around a slowly rotating BH (in Kerr metric) of mass $M=1$ and spin $a=0.1$, having a luminosity parameter $A=0.1$ with a photon impact parameter $b=1$. The test particle motion ends on the critical radius $r_{\rm crit}=2.02M$ (dashed red line), very close to the event horizon $r^+_{\rm (EH)}\equiv1+\sqrt{1-a^2}=1.99M$ (continuous green line), where it starts corotating with constant velocity around the BH, due to the frame dragging effect and the radiation field (see \cite{Bini2009,Bini2011,Defalco20183D}, for details). 

To gain further information on the test body dynamics and the involved radiation processes, we analyse the Rayleigh potential in terms of different variables. Panels b)--f) must be read from bottom up. In panel b), we note that the Rayleigh potential increases almost exponentially with respect to the radial coordinate $r$, starting from the position $r_0=10M$ until the final destination $r=r_{\rm (crit)}$. In panel c), we analyse the Rayleigh potential through the azimuthal coordinate $\varphi$. The initial azimuthal position is $\cos\varphi_0=1$ (i.e., $\varphi_0=0$). Counting how many times the potential profile comes back to the initial position, we can deduce the number of windings, $n_{\rm wind}$, around the BH. In our example, $n_{\rm wind}$ amounts to $1$, as can be easily checked from panel a). The motion ends when the potential reaches its maximum (red dashed line), where the test particle moves with constant velocity on the critical region without changing the value of its Rayleigh potential. 

In Fig. \ref{fig:Fig1}d, we study the behaviour of the Rayleigh potential with respect to the time coordinate $t$. The profile assumes a distinctive $S$-shape, passing from the initial minimum value $V\sim-6.77$ to its maximum $V\sim-1.43$ with a jump in time of $t_{\rm jump}\sim800M$. In order to help the reader figure out how small the latter value is, we calculate four time estimations regarding some relevant astrophysical situations: for the smallest and lightest stellar BH ever observed, knwon as XTE J1650--500 (having mass $M=3.8M_\odot$ \cite{Shaposhnikov2009}), we see that the jump lasts $t_{\rm jump}\sim15.02$ ms; instead for the heaviest stellar BH, GW150914 (with mass $M=62M_\odot$ \cite{Abbott2016}), we obtain $t_{\rm jump}\sim0.25$ s; for an intermediate BH, like the the one recently discovered at the center of 47 Tucanae (having a mass of $M=2300M_\odot$ \cite{Kiziltan2017}), the jump results to be $t_{\rm jump}\sim9.09$ s; finally, for a supermassive BH, like SgrA* in the center of our own Galaxy (having a mass of $M=4\times10^6M_\odot$ \cite{Gillessen2017}), we evaluate a $t_{\rm jump}\sim4.39$ h. 

In panel e), we analyse the Rayleigh potential in terms of the radial velocity, i.e., $\dot{r}\equiv dr/dt$. The graph, possessing vanishing radial velocity both at its starting-point and end-point, assumes a quasi-parabolic shape and has the dashed blue line at $V\sim-3.34$ as quasi-symmetric axis. This profile expresses the fact that the test particle starts decelerating until it reaches the value $\dot{r}\sim-0.13$ (occurring, as can be inferred from plots b) and c), at the position $(r,\varphi)\approx(3M,0.28)$), where it accelerates before smoothly braking on the critical region.  Last panel f) shows the behaviour of the Rayleigh potential with respect to the azimuthal velocity $r\dot{\varphi}\equiv rd\varphi/dt$. The azimuthal velocity is initially zero, afterward it gets its maximum value $r\dot{\varphi}\sim0.37$, then it tends to decrease, and finally to increase again (due to the frame dragging effect), until it attains the constant value $r\dot{\varphi}\sim0.24$.   

In Fig. \ref{fig:Fig2}a we plot the motion of a test particle orbiting a static BH of mass $M=1$ (described in the Schwarzschild metric), affected by luminosity $A=0.3$ with photon impact parameter $b=2$. Like before, the test particle spiral motion ends on the critical region $r_{\rm crit}=2.16M$ (red dashed line). Panels b)--f) must be read from bottom up. We note the following similarities with the former case: in panel b) we can appreciate the typical quasi-exponential grow of the $V$ potential; plot c) expresses the winding of the test particle ($n_{\rm wind}=2$ in this example); panel d) returns again the characteristic $S$-shape, where the jump occurs at $t_{\rm jump}\sim700M$; plot e) exhibits the maximum deceleration line for the radial velocity (dashed blue line) at $\dot{r}=-0.1$; graph f) demonstates how the test particle acquires the maximum azimuthal velocity $r\dot{\varphi}\sim0.32$ before drifting down to the critical radius without increasing its velocity, because there is no frame dragging effect.

Figure \ref{fig:Fig3} refers to the last case analysed. It deviates considerably from the previous two. Indeed in Fig. \ref{fig:Fig3}a, the test particle moves around a rotating BH of mass $M=1$ and spin $a=0.8$ (extreme regime in Kerr metric), endowed with an intense luminosity $A=0.8$ and a photon impact parameter $b=5$. Dynamical motion terminates at spatial infinity, i.e., no critical radius appears. The initial amount of energy suffices to let the test body escape from the two combined attracting force, i.e., the gravitational pull and the PR drag force. In this case, the Rayleigh potential exhibits new features, not encountered before (panels b)--f) must be read from top down). Indeed, in panel b), the Rayleigh potential decreases quasi-exponentially, while in c) it shows a decreasing-monotone behaviour, without winding up around the BH, i.e., $n_{\rm wind}=0$. In graph d), we end up with a reversed trend, since an almost linear decay-shape arises. Unlike the former cases, the potential begins with a maximum value $V\sim-6.76$ and start decreasing at time $t_{\rm dec}\sim60M$ toward the minimum value $V\sim-14.21$. In panel e), we learn that the test particle increases velocity, due to the weakening of both the gravitational pull and the PR drag force. However, asymptotically it should approach the value $\dot{r}\sim0.23$. Also the azimuthal velocity graph f) differentiates itself from Figs. \ref{fig:Fig1}f and \ref{fig:Fig2}f. Indeed, the test particle initially has $r\dot{\varphi}\sim0.3$, then it slows down until it is asymptotically at the rest.  

The huge difference between Figs. \ref{fig:Fig1} and \ref{fig:Fig2} (i.e., test body approaching the critical region) and Fig. \ref{fig:Fig3} (i.e., test particle going to infinity) relies mainly on the three following evidences: ($i$) the Rayleigh potential profile is negative and changes, assuming peculiar and recognizable features for both the situations; ($ii$) if the test particles spirals inward, the Rayleigh potential is monotone-increasing, otherwise it is monotone-decreasing; ($iii$) depending on the case, as a consequence of the remark ($ii$), the plots b)--f) must be read either from bottom up (spiralling inward) or from up down (getting to infinity).             

We decided to draw the test body dynamics in the equatorial plane only \cite{Bini2009,Bini2011}, because the three-dimensional model gives exactly the same results, except that two more plots, related to the $\theta$-motion (position and velocity), must be considered in this case \cite{Defalco20183D} \footnote{We note that this argument is relevant for Kerr metric only. Indeed, for Schwarzschild spacetime, due to the spherical symmetry propriety, all test particles orbits always lie in a plane, reducing thus, with an opportune change of coordinates, to the equatorial plane case.}. Indeed, considering the motion restricted on a plane only permits to highlight handily the relevant aspects of the Rayleigh potential, which we have discussed above.

The plots presented here assign an enormous value to the Rayleigh potential for its observational features. Indeed, by monitoring the test bodies motion around a rotating/static BH it will be possible to infer useful proprieties on the involved radiation processes and, in particular, to reconstruct the functional form of the Rayleigh potential by means of observational techniques through plots b)--f). This step will in turn allow to derive analytically the related radiation force and deduce crucial proprieties regarding the gravitational field (e.g., BH mass and spin), radiation processes (e.g., radiation intensity and photon impact parameter), and dissipation effects (e.g., the role played by the PR effect and the attitude of the system to be influenced by it).

Vice versa, our analysis can also be employed for theoretical purposes. Indeed, by assigning a different functional form to the Rayleigh potential one can straightforwardly calculate the test particle equations of motion and the related trajectories. In such a way, the theoretical investigation of a wide range of radiation processes, including the possible dissipative phenomena, becomes an easy task. Finally, our model allows to benchmark theoretical and synthetic results with observational data.

\subsubsection{Digression on logarithmic Rayleigh potential}
\label{sec:LOGPOT}
The Rayleigh potential (\ref{eq:potential_explicit}) contains the \emph{logarithmic} term $\ln(\mathbb{E}/E_{\rm p})$. This represents a novel aspect in the literature involving relativistic dissipation in radiation processes. In the framework of potential theory, such function has been adopted in different research fields. The implications of a logarithmic potential in Schr\"odinger \cite{Birula1976} and Klein-Gordon equations \cite{Rosen1969,Bartkowski2008} have been examined in the context of non-relativistic quantum mechanics and quantum field theory, respectively. In addition, this kind of nonlinearity appears naturally in inflation cosmology \cite{Barrow1995}, galactic dynamics models \cite{Valluri2012}, and supersymmetric field theories \cite{Enqvist1998}. Besides, there have been profound developments in polynomial and rational approximation theory \cite{Saff2010}, whereas in measure theory it has been fundamental to solve problems arising in electrostatic and classical gravity \cite{Saff1997}. Finally, a recent application consisted in approximating (through a logarithm) the gravitational potential in the regions close to a Schwarzschild BH to analytically describe the motion of test particles and accretion disk structures \cite{Shakura2018}. 

In our model, the logarithmic function is quenched far from the BH by the factor $1/r^2$, while close to it the logarithm dominates, see Eq. (\ref{eq:potential_explicit}). From Sec. \ref{sec:PC}, we realize that the term $\ln(\mathbb{E}/E_{\rm p})$ can be ascribed to the conservative part of the radiation force, $\mathbb{F}_{\rm C}{}^\alpha$. In the classic limit, we obtain
\begin{equation} \label{eq:FCCL}
\mathbb{F}_{\rm C}{}^\alpha\approx\frac{A}{r^2}(1-\dot{r})[1,1,0,0],
\end{equation}
where we recognize that the azimuthal force is zero, i.e., $\mathbb{F}_{\rm C}{}^\varphi=0$. Instead, the radial force is given by 
\begin{equation} \label{eq:FCCL_rad}
\mathbb{F}_{\rm C}{}^r\approx\frac{A}{r^2}-\frac{A}{r^2}\dot{r},
\end{equation}
where $A/r^2$ represents the radiation pressure and $A\dot{r}/r^2$ is half the radial PR drag force, see Eq. (\ref{eq:limF1}). Since in the classical limit the radiation field is constituted by photons travelling along straight lines, radiation absorption occurs only in radial direction. The test particle energy is less than the incoming photon energy, i.e., $\mathbb{E}\le E_{\rm p}$ (Figs. \ref{fig:Fig1} -- \ref{fig:Fig3} confirms that $\ln(\mathbb{E}/E_{\rm p})$ is everywhere negative). Consequently, Eq. (\ref{particle_energy_1}) configures as an absorption energy describing the interaction between the test particle and the radiation field. Indeed, when the test particle is at rest, $\mathbb{E}$ reaches its maximum value (i.e., $\mathbb{E}=E_{\rm p}$), reflecting the fact that the photon energy is entirely absorbed, whereas as the test particle velocity approaches the speed of light, $\mathbb{E}$ tends to zero, since in this case the photon can not hit the test body. In other words, the faster the test body-target moves, the more photon-bullets are \qm{dissipated}, the absorbed energy strongly depending on the test particle velocity. Therefore, we conclude that the logarithmic function is associated with the \emph{absorption process}. 

Bearing in mind the previous observations, we can figure out that the term $U_\alpha U^\alpha$ appearing in Eq. (\ref{eq:potential_explicit}) and stemming from the non-conservative part $\mathbb{F}_{\rm NC}{}^\alpha$ of the radiation force (see Sec. \ref{sec:PNC}), describes the \emph{re-emission process} in the radial and azimuthal directions. The condition according to which the norm of the test particle 4-velocity is constrained to be $-1$, reflects the fact that re-emission is constant, isotropic, and independent of $\boldsymbol{U}$. In addition, since the time component $U^t$ is always nonzero, re-emission will always be present (at least until absorption occurs). 

All the absorbed radiation is afterward completely re-emitted by the test particle, which thus behaves as an ideal black body in thermal equilibrium. Indeed, in our model, absorption and re-emission are intimately intertwined for two reasons. Firstly, the test particle 4-velocity $U^\alpha$ appears both in $\mathbb{V}_{\rm C}$ and $\mathbb{V}_{\rm NC}$. Moreover, both mechanisms would quit if $U^\alpha$ could be replaced with $k^\alpha$, i.e., if the test particle velocity gets really close to the speed of light. 

The above results agree with the hypotheses according to which both absorption and re-emission configure as PR effect causes. Furthermore, it should be clear that, although it is not possible to separate covariantly radiation pressure from PR drag contributions (as stated in Sec. \ref{sec:int_res}), we can clearly distinguish aborption from re-emission moments anyway.

Eventually, it should be stressed that the Rayleigh potential (\ref{eq:potential_explicit}), besides radiation processes, includes also gravitational effects.

\section{Concluding remarks}
\label{sec:end}
Interaction with the environment represents the main feature which differentiates dissipative systems from conservative ones. The approaches aimed at investigating the former are distinct and can be grouped in three main categories, as outlined in the Introduction, Sec. \ref{sec:intro}. However, such frameworks are unlikely to provide a concrete analytical pattern, which could make more feasible the research of the Rayleigh potential in the context of dissipative inverse problem in the calculus of variations.

Notwithstanding, this is exactly the purpose of the \emph{energy formalism}, which is outlined in all its formal aspects in Sec. \ref{sec:EF} by exploiting the powerful machinery of differential geometry. This ensures the model a general structure and hence a wide applicability. In particular, it turns out to be well suited for any metric theory of gravity. Equations (\ref{eq:constraint}) and (\ref{eq:constraint_2}) constitute the core of the \emph{energy formalism}, guaranteeing a twofold advantage. Firstly, a sensible reduction of the calculations, underlying the determination of the $V$ potential related to an exact differential semi-basic one-form $\boldsymbol{\omega}$ (cf. Eq. (\ref{eq: differential_of_V})), is achieved, since the integration process involves only the energy $\mathbb{E}$, instead of the $n$ variables $\boldsymbol{U}$ (see Eq. (\ref{eq:pot_E})). In addition, the obtained expression of the $V$ potential, as a function of $\mathbb{E}$, suffices for the description of the dynamics. This represents a crucial point, whenever the evaluation of $f(\boldsymbol{X},\boldsymbol{U})$ turns out to be too laborious (see discussion following Eqs. (\ref{eq:pot_E}) and (\ref{eq:derivata_di_f})). In this regard, \emph{energy formalism} shares strong similarities with the approach devised by Lagrange and D'Alembert for conservative dynamical systems.

In order to appreciate the formal apparatus introduced in the previous section, in Sec. \ref{sec:TPE} we have proposed the simple model of a dynamical system admitting a Rayleigh potential which is cubic in the velocity variable. This example grants a twofold advantage. On the one hand, it demonstrates that dissipative forces which are linear, quadratic or more in general polynomial with respect to the velocity field can be easily integrated both through the standard integration method and the \emph{energy formalism}. On the other hand, it makes clear that criticalities naturally arise when dissipative forces depending non-linearly on the velocity field are considered, which are customary in gravitational contexts. 

In Sec. \ref{sec:AEF}, we have applied the \emph{energy formalism} to the general relativistic PR effect, obtaining for the first time in the literature the analytic form of the Rayleigh potential, Eq. (\ref{eq:potential_explicit}). To this aim, the role played by the integrating factor (\ref{eq:if2}) has proved to be crucial. 

The Rayleigh potential fulfils an important role in many modern research fields. In particular, recent studies involving contact manifolds have shown how such potential allows a considerable simplification of Lagrangian problems with friction \cite{Minguzzi2015}; further developments  both in classical and quantum settings can be found in Ref. \cite{Ciaglia2018}. In addition, in this paper we have proposed new original  applications, which can be summarized as follows (see Sec. \ref{sec:int_res} and Ref. \cite{DBletter2019}, for more details):
\begin{itemize}
    
    \item the most significant contributions are displayed in Figs. \ref{fig:Fig1}--\ref{fig:Fig3}, where we have described how to link coherently the observations with the theoretical results and viceversa. This allows to acquire useful information on the mathematical structure and on the physical properties of the phenomenon analysed. These results might confer a prominent observational relevance to the PR model;
    
    \item a new functional class, represented by the logarithm function occurring in Eq. (\ref{eq:potential_explicit}), has been discovered. At the best of our knowledge, this is a new facet in the literature. We have ascertained that such term describes absorption processes. This may entail significant implications in the context of both PR effect and, more in general, radiation processes in high-energy astrophysics.
    
    The last point allows us to stress a crucial aspect of the \emph{energy formalism}. Indeed, if we were not able to determine the analytic expression of $f(\boldsymbol{X},\boldsymbol{U})$, anyway we would have obtained the logarithmic function and hence the fundamental proprieties of the absorption mechanism (as we actually did). This in turn would have allowed, in any case, a complete description also of the emission process, enclosed in the $f$ function. Indeed, by exploiting the classical limit and qualitative arguments (based on the precise knowledge of the logarithmic factor), it would have been possible to deduce the essential features or even the functional form of $f$; 
    
   \item the combined use of the integrating factor and the \emph{energy formalism} gives rise to a novel instrument to determine analytically the Rayleigh dissipation function in non-linear dynamical problems framed in gravitational settings.  
\end{itemize}

As part of our future research program we intend to exploit \emph{energy formalism} in the context of gravitational waves.

\section*{Acknowledgements}
The authors are grateful to Professor Luigi Stella, Doctor Giampiero Esposito, Professor Giuseppe Marmo, and Professor Tom Mestdag for the stimulating discussions and the useful suggestions aimed at improving the scientific impact of this formalism. The authors are grateful to Professor Antonio Romano for the useful discussions. The authors thank the Silesian University in Opava and the International Space Science Institute in Bern for hospitality and support. The authors are grateful to Gruppo Nazionale di Fisica Matematica of Istituto Nazionale di Alta Matematica for support.

\bibliography{references}
\end{document}